\documentclass{aa}
\usepackage[varg]{txfonts}

\usepackage[utf8]{inputenc}
\usepackage[export]{adjustbox}
\usepackage{siunitx}
\usepackage{amssymb}
\usepackage{amsmath}
\usepackage{mathtools}
\usepackage{multirow}
\usepackage{relsize}
\usepackage{soul}
\usepackage{xcolor}

\usepackage[switch,mathlines]{lineno} 

\usepackage{natbib}
\bibpunct{(}{)}{;}{a}{}{,} 

\usepackage[colorlinks=true,linkcolor=red,citecolor=black]{hyperref}

\makeatletter
\renewcommand*\aa@pageof{, page \thepage{} of \pageref*{LastPage}}
\makeatother

\begin{document} 


\title{
Detectability of large-scale counter-rotating stellar disks in galaxies with integral-field spectroscopy
}
\author{
M. Rubino \inst{1}\thanks{\email{michela.rubino@phd.unipd.it}}
\and 
A. Pizzella\inst{1,2}
\and 
L. Morelli\inst{3}
\and 
L. Coccato\inst{4}
\and 
E. Portaluri\inst{5,6}
\and 
V. P. Debattista\inst{7}
\and 
E. M. Corsini\inst{1,2}
\and
\mbox{E. Dalla Bont\`{a}\inst{1,2}}
}

\institute{
Dipartimento di Fisica e Astronomia ``G. Galilei'', Universit\`a degli Studi Padova, vicolo dell'Osservatorio 3, I-35122 Padova, Italy
\and INAF - Osservatorio Astronomico di Padova, vicolo dell'Osservatorio 5, I-35122 Padova, Italy
\and Instituto de Astronom\`{i}a y Ciencias Planetarias, Universidad de Atacama, Copayapu 485, Copiap\'{o}, Chile
\and European Southern Observatory, Karl-Schwarzschild-Strasse 2, D-85748 Garching, Germany \and 
INAF - Osservatorio Astronomico d'Abruzzo, Via Mentore Maggini, I-64100 Teramo, Italy
\and ADONI - Laboratorio Nazionale di Ottica Adattiva, Italy
\and Jeremiah Horrocks Institute, University of Central Lancashire, Preston, PR1 2HE, UK
}
     
\date{Received date/ accepted date}

\abstract
{In recent years integral-field spectroscopic surveys have revealed that the presence of kinematically decoupled stellar components is not a rare phenomenon in nearby galaxies. However, complete statistics are still lacking because they depend on the detection limit of these objects.}
{We investigate the kinematic signatures of two large-scale
counter-rotating stellar disks in mock integral-field spectroscopic data to address their detection limits as a function of the galaxy properties and instrumental setup.}
{We built a set of mock data of two
large-scale counter-rotating stellar disks as if they were observed with the Multi Unit Spectroscopic Explorer (MUSE). We accounted for different photometric, kinematic, and stellar
population properties of the two counter-rotating components as a function of galaxy inclination. 
We extracted the stellar kinematics in the wavelength region of the calcium triplet absorption lines by adopting a Gauss-Hermite (GH)
 parameterization of the line-of-sight velocity distribution (LOSVD).}
{We confirm that the strongest signature of the presence of two counter-rotating stellar disks is the symmetric double peak in the
velocity dispersion map, already known as the $2\sigma$ feature. 
The size, shape, and slope of the 2$\sigma$ peak strongly depend on the velocity separation and relative light contribution of the two counter-rotating stellar disks. When the $2\sigma$ peak is difficult to detect due to the low signal-to-noise ratio of the data, the large-scale structure in the $h_3$ map can be used as a diagnostic for strong and weak counter-rotation. 
The counter-rotating kinematic signatures become fainter at lower
viewing angles as an effect of the smaller projected velocity separation between the two counter-rotating components. 
We confirm that the observed frequency of $2\sigma$ galaxies represents only a lower limit of the stellar counter-rotation phenomenon.}
{The parameterization with a single GH function does not provide a good description of the LOSVD in the presence of strong counter-rotation. However, using   GH
parametric solutions is a practical way to reveal the large-scale counter-rotating
stellar disks and could be used to detect faint counter-rotating components to improve the statistics of stellar counter-rotation.}



\titlerunning{Detectability of counter-rotating stellar disks}
\authorrunning{M. Rubino et al.}

\maketitle

\section{Introduction}\label{sec:intro}

Galaxies experience several star formation episodes and interaction events during their lifetimes, resulting in complex systems that might have significantly changed their morphological appearance, orbital structure, and stellar population properties. 
All the different components that a galaxy might contain (such as bulges, disks, bars, shells) are indeed the result of these multiple episodes. Therefore, understanding how a galaxy formed and built up its various components is challenging. 
One key tool for investigating the formation processes is   to disentangle the properties of the stellar populations (kinematics, age, metallicity, chemical abundances, and luminosity)  in each structural component \citep[e.g.][]{corsini2016YoungNuclearStellar,morelliStellarPopulationsBulges2016,zhuMorphologyKinematicsOrbital2018,tabor2019SDSSIVMaNGA,ohSAMIGalaxySurvey2020,zhuDisentanglingFormationHistory2020}. 

In this respect, galaxies hosting two or more counter-rotating components offer a unique astrophysical laboratory to study how some galaxy components formed and evolved.
A stellar and/or gaseous counter-rotating component is characterized by an opposite angular momentum with respect to the main stellar body of the host galaxy.
This results in peculiar kinematic signatures that allow us to separate the relative contribution of the different components that are counter-rotating, and to independently study their spectral properties \citep[see][for reviews]{gallettaCounterrotationBarredGalaxies1996, bertolaCounterrotationGalaxies1999, corsiniCounterRotationDiskGalaxies2014}.

Large-scale counter-rotating stellar disks are of particular interest among the several kinds of counter-rotation phenomena. 
The first detection of two large-scale counter-rotating stellar disks was done by \cite{rubinCospatialCounterrotatingStellar1992} by studying the shape of the absorption lines in the long-slit spectra of the E7/S0 galaxy NGC~4550. 
Its counter-rotating disks are co-spatial and have similar sizes, luminosities, and masses \citep{rixNGC4550Laboratory1992, johnstonDisentanglingStellarPopulations2013}, but different thickness \citep{cappellariSAURONProjectOrbital2007}. 
Moreover, one of them is associated with a disk of ionized gas \citep{rubinCospatialCounterrotatingStellar1992,sarziSAURONProjectIntegralfield2006,coccatoSpectroscopicEvidenceDistinct2013}. 
In addition to NGC~4550, a number of galaxies hosting two counter-rotating stellar disks with different sizes, luminosities, and masses are also known \citep[e.g.,][]{bertolaCounterrotatingStellarDisks1996}. 

Counter-rotating stellar disks are more easily detected in the outermost regions of galaxies where the seeing blurring is negligible, the contamination of bulge and bar is not relevant, and the velocity separation between the counter-rotating components is maximum giving rise to double-peaked absorption-line spectra. The line-of-sight velocity distribution (LOSVD) turns out to be bimodal, and each of its two parts corresponds to one of the two counter-rotating stellar disks. 
Several techniques have been developed to perform a spectroscopic decomposition and derive the relative contribution of the counter-rotating components to the measured LOSVD as a function of position onto the sky-plane. 
In this way it is possible to recover their spatial distribution, kinematics, and stellar population properties \citep[][]{coccatoDatingFormationCounterrotating2011,katkovStarsIonizedGas2011, johnstonDisentanglingStellarPopulations2013, mitzkusDominantDarkMatter2017,mendez-abreuSpectrophotometricDecompositionGalaxy2019,barrosoLOSVD2021}. 
All these methods suffer severe limitations when the velocity separation and/or light contribution of the two counter-rotating stellar components are small. 
However, when the spectroscopic decomposition is successful it is possible to constrain the formation timescale of the counter-rotating components and recover the assembly process of the host galaxy \citep[e.g.,][]{pizzellaDifferenceAgeTwo2014, morelliKinematicStellarPopulation2017, nedelchevPropertiesKinematicallyDistinct2019}.

Different scenarios involving an external or internal origin were proposed to explain the formation of a counter-rotating stellar disk. 
It could be the end result of star formation in a counter-rotating gaseous disk, which was accreted through episodic or continuous acquisition of gas clouds from the environment \citep{thakarFormationMassiveCounterrotating1996, thakarSmoothedParticleHydrodynamics1998} or through the capture of a gas-rich satellite during a minor merger \citep{thakarNGC4138Case1997, bassettFormationS0Galaxies2017}. 
Major mergers can form two counter-rotating stellar disks of similar size and mass, but only under some very specific initial conditions \citep{puerariFormationMassiveCounterrotating2001,crockerMolecularGasStar2009,martelFormationCounterrotatingStars2020}.
Another viable external formation process was revealed by cosmological simulations when the effects of the torque of two distinct filamentary structures triggered the star formation onto both prograde and retrograde orbits \citep{algorryCounterrotatingStarsSimulated2014}.
In addition, \cite{kanthariaTorquesAngularMomentum2016} suggested that the mutual gravity torques between nearby galaxies can play an important role in changing the disk angular momentum and stressed the importance of environment in the assembly of counter-rotating components.
An alternative to the external origin is the internal dynamical instability of the system itself that can lead to the dissolution of a bar or triaxial stellar halo with the formation of a counter-rotating component \citep{evansSeparatrixCrossingEnigma1994}.
However, the internal origin does not account for the observed properties of all the known counter-rotating galaxies, which are better explained by mergers and interactions, although the details of this building process are not   fully understood yet \citep{corsiniCounterRotationDiskGalaxies2014}. 
To this end, we first need to precisely address the demography of counter-rotating stellar disks to determine  whether they are a widespread phenomenon in lenticulars and spirals or a class of rare, although fascinating, structures. 

Early studies of the frequency of counter-rotating disks of gas and/or stars in lenticulars and spirals were mostly based on long-slit spectroscopic observations of limited samples.
The frequency of counter-rotating gaseous disks in lenticulars is $\sim30\%$ \citep[][]{pizzellaIonizedGasStellar2004}. 
This reveals that gas accretion events are not rare in S0 galaxies.
On the other hand, only $\sim10\%$ of them host a significant fraction ($>5\%$) of counter-rotating stars \citep[][]{kuijkenSearchCounterRotatingStars1996}. 
\cite{pizzellaIonizedGasStellar2004} found that $<12\%$ and $<8\%$ of spirals host a gaseous or a stellar counter-rotating disk, respectively. 
They explained the lower frequency of counter-rotating spirals with respect to counter-rotating lenticulars by the requirement of a larger amount of external gas to first form a counter-rotating gaseous disk that subsequently turns into a counter-rotating stellar disk. 
Spirals are indeed gas-rich systems and the acquired retrograde gas is swept away if it is less than the pre-exiting prograde one.
On the contrary, gas-poor lenticulars can build a counter-rotating gaseous disk even if a small amount of external retrograde gas is acquired.

A step forward in disentangling the different kinematic components of galaxies and in revealing hidden structures like counter-rotating stellar disks has been made possible by integral-field unit (IFU) spectroscopic observations \citep[e.g.,][]{krajnovicATLAS3DProjectII2011,jinSDSSIVMaNGAProperties2016,bryantSAMIGalaxySurvey2019}.
The analysis of large samples of galaxies gives important hints about their formation process \citep[e.g.,][]{liImpactMergingOrigin2021}, and provides criteria for their classification \citep[e.g.,][]{vandesandeSAMIGalaxySurvey2017}.
\citet{krajnovicATLAS3DProjectII2011} separated the 260 early-type galaxies of the ATLAS$^{\rm 3D}$ survey \citep{cappellariATLAS3DProjectVolumelimited2011} into five groups according to their stellar kinematic maps. 
The so-called double-sigma ($2\sigma$) galaxies are characterized by two offset and symmetric peaks of the stellar velocity dispersion along the galaxy major axis. 
This feature was already identified as the signature of the presence of two counter-rotating disks co-existing in the host galaxy from long-slit data \citep[e.g.,][]{rixNGC4550Laboratory1992,bertolaCounterrotatingStellarDisks1996,verganiNGC571913Interacting2007} and used as a diagnostic in the search for counter-rotating galaxies.
\citet{krajnovicATLAS3DProjectII2011} found that the frequency of $2\sigma$ galaxies is $4\%$ when considering this as a lower limit because most of them are lenticular galaxies seen at high inclinations. 
Moreover, the $2\sigma$ signature in some objects could be undetected depending on the instrumental velocity dispersion and the properties of the counter-rotating stellar disks. 
Though the $2\sigma$ feature is the compelling evidence of counter-rotating stellar disks, a dedicated study is still necessary to investigate the observational limits for their detection in stellar kinematic maps.

The detection of a counter-rotating component depends on either instrumental (e.g., spectral and spatial resolution) and physical characteristics of both prograde and counter-rotating components. 
At present, the most suitable instrument to search for counter-rotating signatures is the {\em Multi Unit Spectroscopic Explorer} (MUSE).
In particular, this instrument covers a wide wavelength range that includes the calcium triplet (CaT) absorption lines ($\lambda\lambda8498, 8542, 8662$~\AA) and its instrumental velocity resolution at $\lambda=8600~\text{\AA}$ is $\sigma_{\rm MUSE} = 38$ km~s$^{-1}$. 
This resolution allows us to better resolve the absorption lines and obtain more precise kinematic measurements.
Comparing the MUSE instrument to other IFU spectrographs used for galaxy surveys, such as ATLAS$^{\rm 3D}$ \citep{cappellariATLAS3DProjectVolumelimited2011}, CALIFA \citep{sanchezCALIFACalarAlto2012}, SAMI \citep{bryantSAMIGalaxySurvey2015}, and MaNGA \citep{bundyOverviewSDSSIVMaNGA2015}, only the last is found to  cover a wider spectral range, but its instrumental dispersion is larger than $50$ km~s$^{-1}$ and its spatial resolution is $\sim2$ arcsec. The kinematic measurements of MaNGA (and SAMI) are severely impacted by atmospheric seeing \citep{vandesandeSAMIGalaxySurvey2020}.

This work consists of a detailed study of the stellar kinematic maps of mock counter-rotating galaxies observed with MUSE to define new useful diagnostics for identifying the presence of two large-scale counter-rotating stellar disks in real galaxies, and hence addressing their statistics. 
We want to both investigate hidden and/or unclear kinematic signatures and unveil faint and/or unresolved counter-rotating components. 
In particular, we focus on those cases where the small velocity separation, similar velocity dispersions, similar stellar populations, or  different light contributions make it difficult to detect and analyze the counter-rotating components. 

The paper is organized in the following way. In Sect.~\ref{sec:LOSVD} we explore how the presence of a counter-rotating component shapes the spectrum of a prograde and brighter component, and investigate the goodness of the recovery of the LOSVD.
In Sect.~\ref{sec:approach} we extend our analysis to mock IFU data representing two counter-rotating stellar disks in order to highlight the counter-rotating signatures in the observed kinematic maps. 
In Sect.~\ref{analysis} we present the analysis of the stellar kinematics.
In Sect.~\ref{sec:limits} we discuss some possible limitations to our approach, and   we conclude in Sect.~\ref{sec:CONC}.

\section{The LOSVD of two counter-rotating components}\label{sec:LOSVD} 
To search for counter-rotating signatures in the stellar kinematics we need to investigate the possible shapes of the LOSVDs that differ not only because of prograde and retrograde motions but also on account of the relative light contribution of the two decoupled components.

We assume that each component is represented by an E-MILES single stellar population (SSP) model \citep[][]{vazdekisEvolutionaryStellarPopulation2010,vazdekisEvolutionaryStellarPopulation2015} to be convolved with different Gaussian LOSVDs. 
These models are characterized by spectral resolution FWHM$_{\rm ssp} = 2.51$~\AA\ and spectral sampling ($\Delta \lambda$)$_{\rm ssp} = 0.9$ \AA~pixel$^{-1}$, and cover a wide spectral range (1680--50000 \AA) including the CaT triplet absorption lines. 
The SSP models are based on BaSTI isochrones \citep{pietrinferniLargeStellarEvolution2004} and bimodal initial mass function (IMF) with slope of 1.30. 
They span a wide range of ages ($0.03\leq t \leq14$ Gyr) and metallicities ($-2.27\leq [\rm{M}/\rm{H}] \leq 0.40$ dex).
These models do not explicitly consider any $\alpha$-enhancement and assume that $[\rm{M}/\rm{H}]=[\rm{Fe}/\rm{H}]$. 

For our analysis we decided to adopt a model with age of $t=6$ Gyr and metallicity $[\rm{M}/\rm{H}]=-0.25$ dex. 
These values are consistent with those measured at 1$r_{\rm e}$ by \citet{kuntschnerSAURONProjectXVII2010} for the most widely studied counter-rotating galaxy NGC~4550. 
On the other hand, we do not expect our analysis to change significantly adopting different ages and metallicities.
Stellar population properties mainly affect the continuum and equivalent width of the spectral lines and have a secondary effect on their positions. Using a single age and metallicity allows us
to limit the number of free parameters and highlights the effects induced by the kinematics.
For the two components we used the same SSP model, namely the same age and metallicity, to avoid biasing the fractional light contribution due to intrinsic differences (e.g., young and old stars have different spectral energy distributions) when combining the two templates.
However, our  aim was to model and measure the relative contribution of the counter-rotating component with respect to the prograde component (or equivalently to the total galaxy) without any attempt to recover the input properties of the stellar populations. Therefore, the choice of the stellar parameters is not relevant because they are equal for both components.

For the prograde component, we fix the relative light contribution to unity and let the other vary considering a multiplicative factor $f$ ranging from 0.1 to 1.0 with a step of 0.1, such that the light contribution of the counter-rotating component with respect to the total galaxy $f_{\rm cr}$ varies from 0.09 to 0.50. 
The resulting mock spectrum of the counter-rotating galaxy is given by 
\begin{equation}
S_{\rm m}(\lambda) =T({\lambda}) * G(\lambda|V_{\rm pro},\sigma_{\rm pro})+f \times T({\lambda}) * G(\lambda|V_{\rm cr},\sigma_{\rm cr})~,
\end{equation}
where $\lambda$ is the wavelength, $T(\lambda)$ is the stellar template and $G(\lambda|V,\sigma)$ is the Gaussian LOSVD characterized by mean velocity $V$ and velocity dispersion $\sigma$. The symbol $*$ denotes convolution.

We built a set of mock galaxy spectra in the wavelength range between 8000 \AA\/ and 9000 \AA\/ that is centered on the CaT triplet. 
For the convolution, we define the mean velocity and velocity dispersion of the prograde ($V_{\rm pro},\sigma_{\rm pro}$) and counter-rotating component ($V_{\rm cr},\sigma_{\rm cr}$).
We choose $V_{\rm cr} = -V_{\rm pro}$ such that the velocity separation $\Delta V$ between the two components spans the range from 5 km~s$^{-1}$ to 400 km~s$^{-1}$ with a step of 5 km~s$^{-1}$. 
This matches the maximum amplitude of stellar rotation curves, which typically does not exceed 200 km~s$^{-1}$. 
For the velocity dispersion, we fix the value of the prograde component $\sigma_{\rm pro}=50$ km~s$^{-1}$ and vary $\sigma_{\rm cr}$ from 40 km s$^{-1}$ to 100 km s$^{-1}$ with a step of 5 km s$^{-1}$. 
These values are consistent with those found in NGC~4550 \citep[e.g.,][]{coccatoSpectroscopicEvidenceDistinct2013}.
With this choice the ratio $\sigma_{\rm cr}/\sigma_{\rm pro}$ covers the interval from 0.8 to 2.0. 
The range of the parameters of the mock spectra is given in Table~\ref{tab:input}.
\begin{table}
    \caption{Initial and final values and step size for the parameters of the mock spectra.}
    \label{tab:input}
    \centering
    \centering
    \begin{tabular}{cccc}
\hline \hline
\rule{0pt}{2.5ex}
 Parameter & Start & End & Step \\
(1) & (2) & (3) & (4) \\
\hline
\rule{0pt}{2.5ex}
$V_{\rm pro}~[{\rm km~s^{-1}}]$ & 2.5  & 200.0  & 2.5 \\
$\Delta V~[{\rm km~s^{-1}}]$ & 5.0  & 400.0  & 5.0\\
$\sigma_{\rm cr}~[{\rm km~s^{-1}}]$ & 40.0  & 100.0  & 5.0\\
$\sigma_{\rm cr}/\sigma_{\rm pro}$ & 0.8  & 2.0  & 0.1\\
$f$ &  0.1 & 1.0  & 0.1\\
$f_{\rm cr}$ & 0.09  & 0.50  & -- \\
\hline 
\rule{0pt}{1ex}
\end{tabular}\\
\begin{minipage}{\textwidth}\small
{\bf Notes.} The $f_{\rm cr}$ parameter is derived from $f$ and therefore has no step.
\end{minipage}  
\end{table}

We assume that the velocity and velocity dispersion are the observed quantities along the LOS.
Moreover, to mimic observations, we consider an instrumental broadening function and a spectral sampling matching those of MUSE (i.e., FWHM$_{\rm MUSE}=2.54$ \AA\ and \mbox{$(\Delta \lambda)_{\rm MUSE} = 1.25$ \AA\ pixel$^{-1}$}).
Whereas, we do not add detector, sky, and photon noise to the mock spectra, which is equivalent to having very high signal-to-noise ratio ($S/N$) data. 
We can account for noise a posteriori only if we require it.

The absorption lines of the mock spectra appear broadened, asymmetric, or even double-peaked depending on  the relative light contribution and the velocity separation of the two counter-rotating components. 
This demonstrates the large variety of possible LOSVD shapes we have to deal with in the case of counter-rotating stellar systems. 
Since it is not straightforward to quantify resolved and unresolved stellar counter-rotation, we only define a criterion for the detection of counter-rotating components based on the recovered shape of the LOSVD by using the Gauss-Hermite (GH) parameterization \citep{vandermarelNewMethodIdentification1993,gerhardLineofsightVelocityProfiles1993}.

\subsection{The Gauss-Hermite parameterization}\label{sec:GHparam}
\begin{figure*}
\centering
\includegraphics[width=0.93\textwidth]{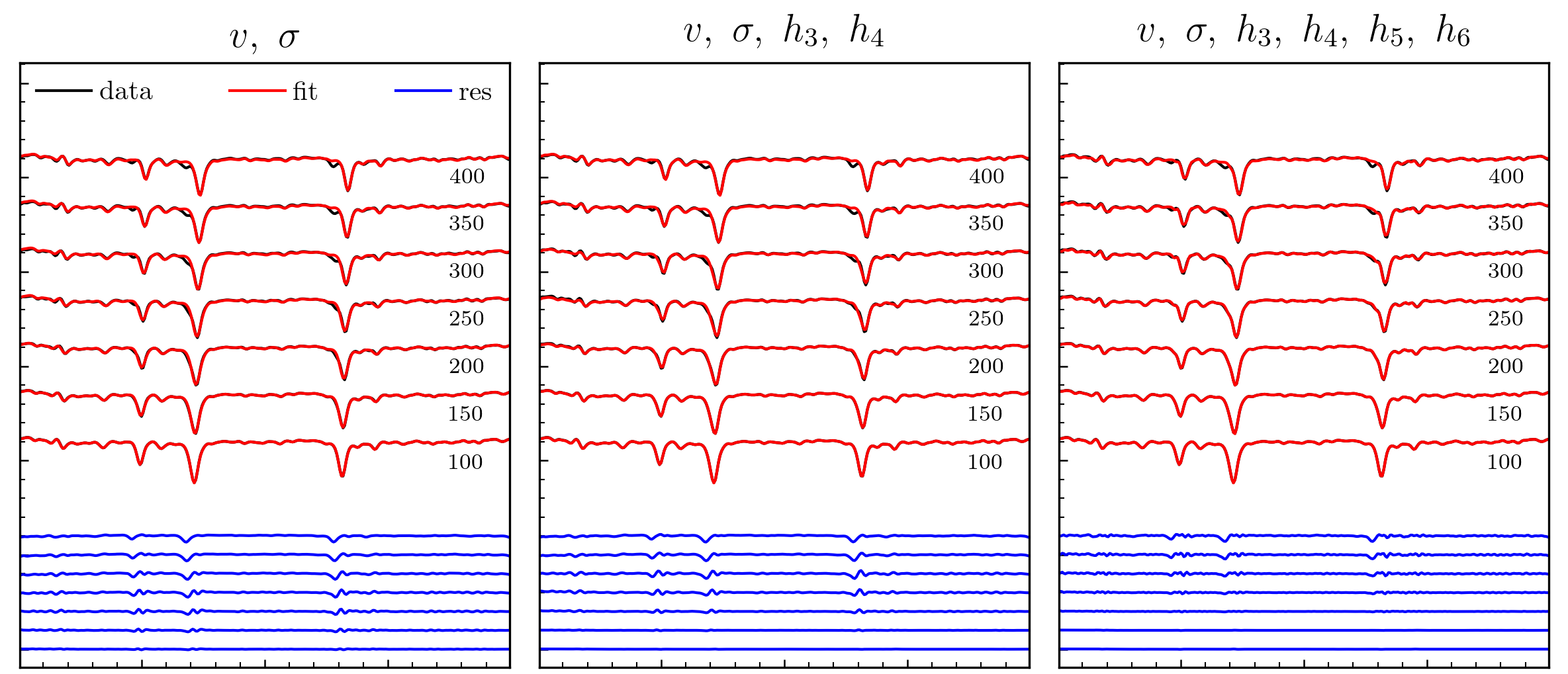}\\
\includegraphics[width=0.93\textwidth]{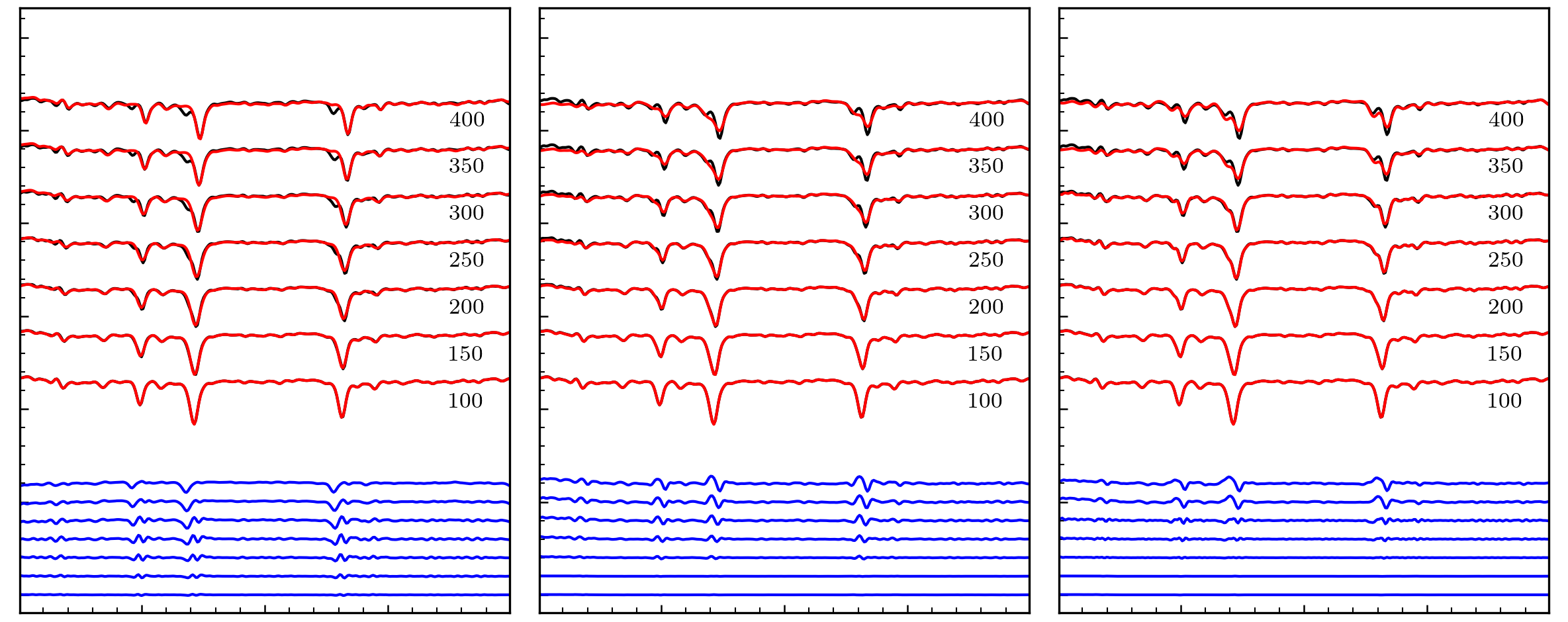}\\
\includegraphics[width=0.93\textwidth]{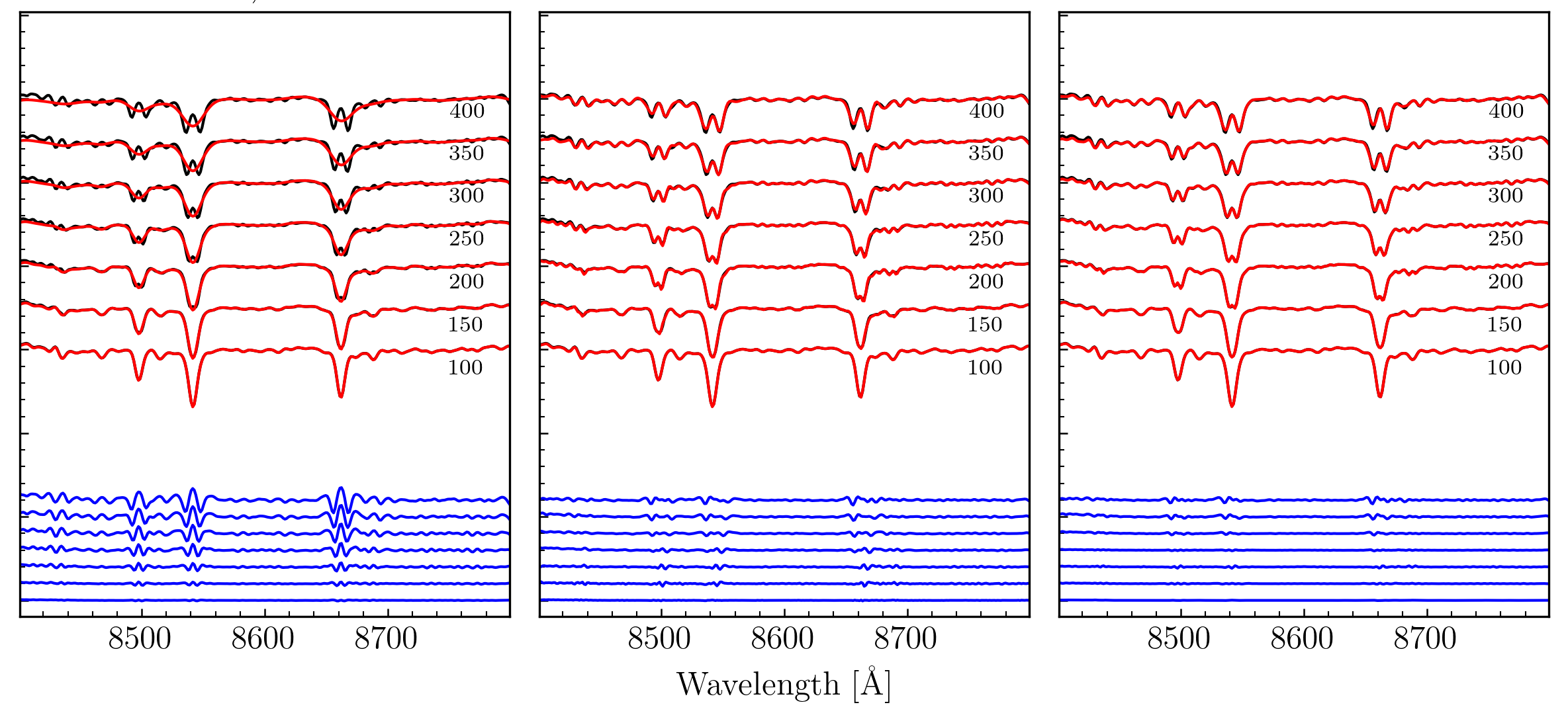}\\
\caption{Best fit ({\em red lines\/}) of some mock spectra ({\em black lines\/}) with their residuals ({\em blue lines\/}) obtained using the GH parameterization for three sets of kinematic measurements: 1) lower-order moments only ($v$, $\sigma$, {\em left panels\/}); 2) four moments ($v,\sigma,h_3$, $h_4$, {\em middle panels\/}); 3) six moments ($v,\sigma,h_3,h_4,h_5$, $h_6$, {\em right panels\/}). 
Each panel shows mock spectra with decreasing velocity separation from top to bottom as labeled, while the constant contribution of the counter-rotating component increases from top to bottom   ($f_{\rm cr}=0.17,0.23$, and $0.50$, respectively). }
\label{fig:mockfit}
\end{figure*}
To understand the counter-rotating signatures we extract the stellar kinematics from our set of mock spectra by using the standard (single) GH parameterization. 
This analysis allows us to investigate the limitations of the recovery of the LOSVD of counter-rotating systems when adopting this commonly used method and, at the same time, to test a possible detection method for counter-rotating components based on the measurement of non-Gaussian absorption lines.

We use the Penalized Pixel-Fitting code \citep[pPXF;][]{cappellariParametricRecoveryLineofSight2004, cappellariImprovingFullSpectrum2017}, which fits the galaxy spectra by convolving a stellar template with the LOSVD modeled as a truncated GH series. 
This implementation uses the Hermite polynomials assuming $h_0,h_1,h_2=1,0,0$.
We adopt as stellar templates a full set of E-MILES SSP models degraded to instrumental resolution.
The best-fit solution is a non-negative linear combination of these templates obtained from a $\chi^2$ minimization in pixel space. 
In particular, we perform three sets of kinematic measurements: 1) considering $v$ and $\sigma$ only; 2) including $h_3$ and $h_4$; 3) truncating the series at the sixth-order term ($V,\sigma,h_3,h_4,h_5,h_6$).
We do not include any additive polynomials, but use a multiplicative polynomial of order four to account for the shape of the continuum.

In Fig.~\ref{fig:mockfit} we show some examples of best-fit solutions for each kinematic set and different $f_{\rm cr}$.
We show our mock spectra with their best fit and corresponding residuals. 
The mock spectra are labeled with the value of the input velocity separation $\Delta V$ and are shifted, as are     the residuals, to avoid overlapping. 
As expected, the residuals decrease with increasing order of the truncation of the GH series and with decreasing $\Delta V$ and $f_{\rm cr}$. 
The fit performed with only the lower-order moments of LOSVD is obviously poor.
In particular, for $f_{\rm cr}=0.17$ (Fig.~\ref{fig:mockfit}, {\em top panels\/}) or less we do not detect the small contribution of the counter-rotating component, even when including in the fit the higher-order moments.
On the contrary, for $f_{\rm cr}=0.23$ (Fig.~\ref{fig:mockfit}, {\em middle panels\/}), using at least $h_3$ and $h_4$, we start detecting the secondary component producing asymmetries of the lines, although the residuals are higher than in the case with $f_{\rm cr}=0.50$ (Fig.~\ref{fig:mockfit}, {\em bottom panels\/}). 
This problem could be solved, for example, by removing the constraints of $h_1 = h_2 = 0$, but this is beyond the scope of this work.

\begin{figure}
\centering
\includegraphics[width=0.5\textwidth]{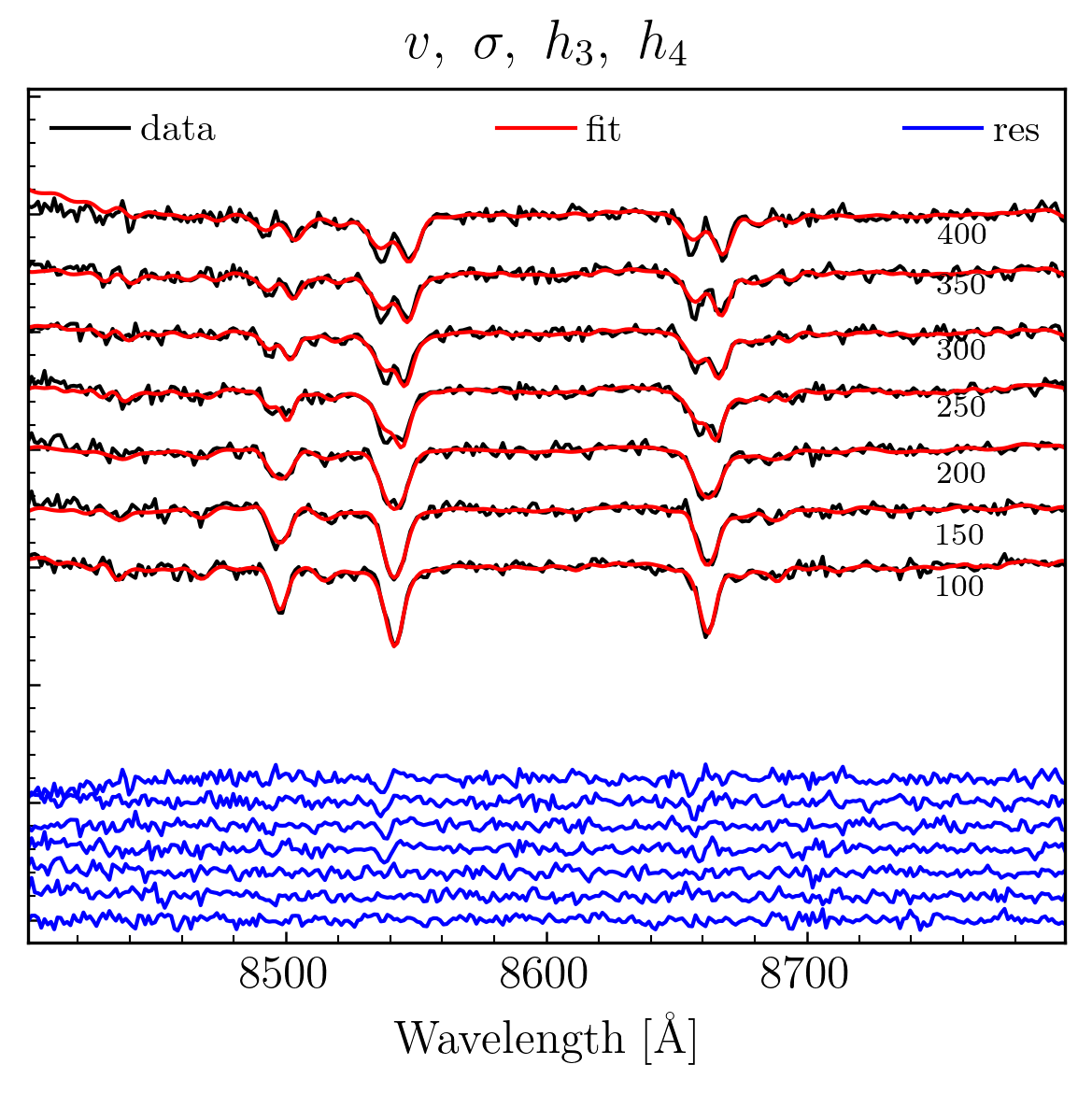}
\caption{As in Fig.~\ref{fig:mockfit}, but   shown are the best fit to mock spectra with $f_{\rm cr}=0.50$ and $S/N=50$ pixel$^{-1}$ obtained using four GH moments.}
\label{fig:mocknoise}
\end{figure}

Despite the goodness of the fit for equal contributing components, in realistic cases we do not have very high $S/N$ data. 
Thus, for comparison we include noise in the spectra of the last example to simulate data with $S/N=50$ pixel$^{-1}$. 
It is not surprising that the noise reduces the ability of recovering small features or double-peaked lines due to the presence of the secondary component (Fig.~\ref{fig:mocknoise}).
However, the broadening of the lines and their strong asymmetries are still recovered as   can be seen from trend of the measured parameters. 
\begin{figure*}
\centering
\includegraphics[width=\textwidth]{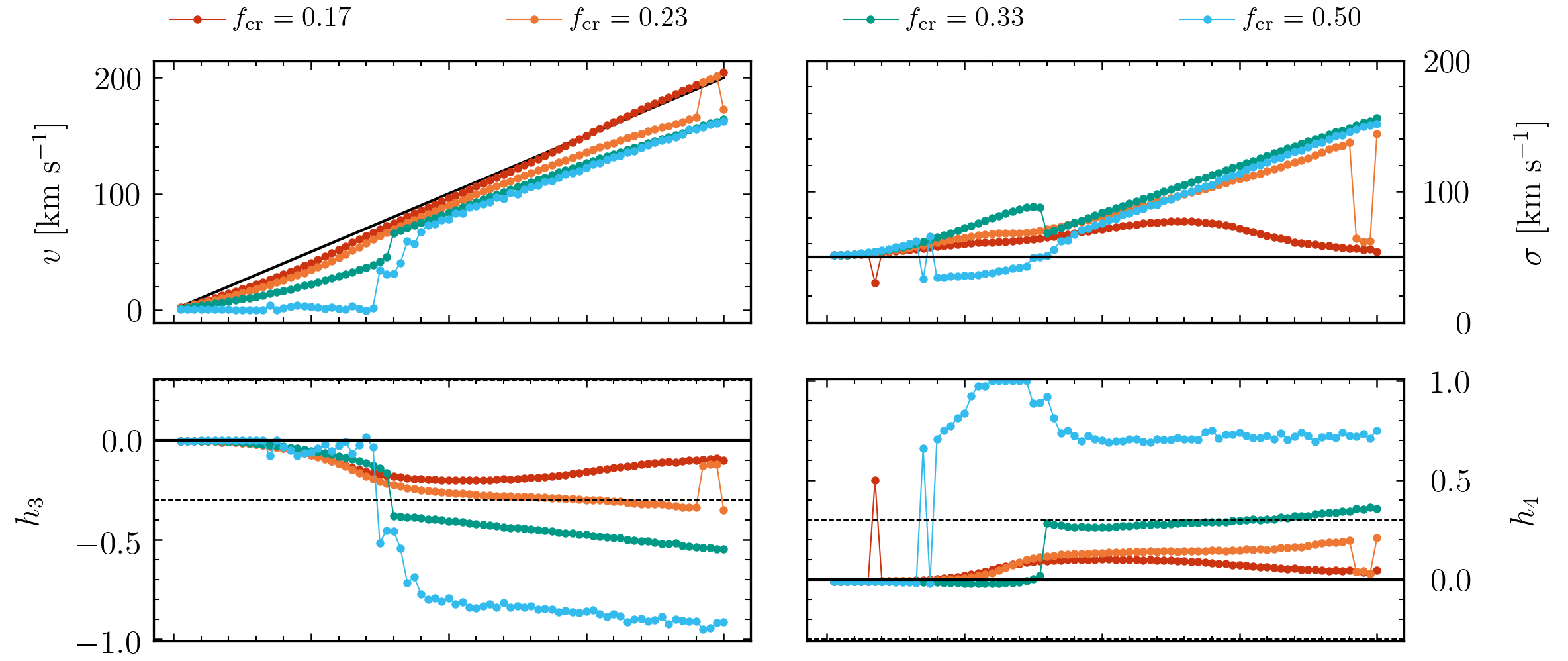}\\
\includegraphics[width=\textwidth]{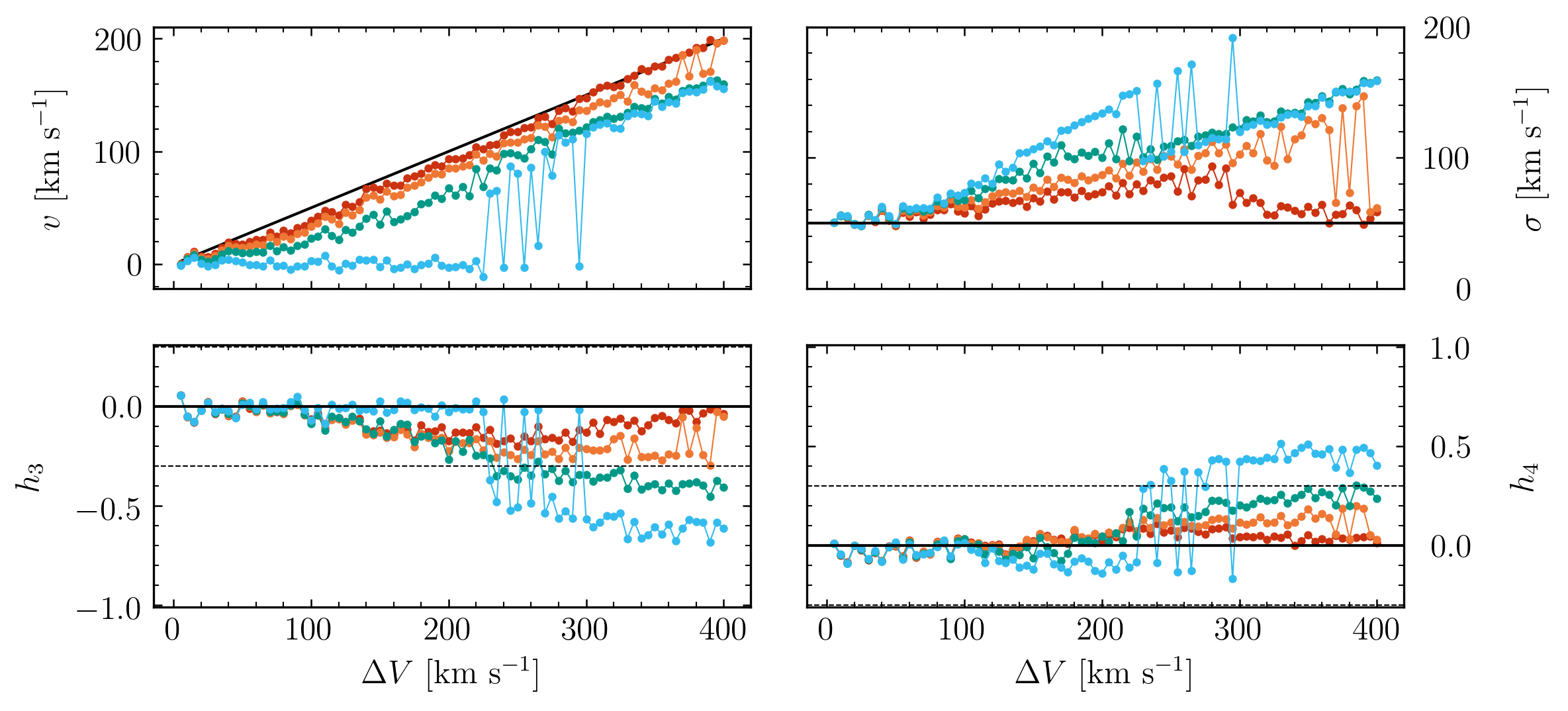}\\
\caption{Comparison of the recovered stellar kinematic parameters in the case of mock spectra without noise ({\em top four panels}) and with $S/N=50$ pixel$^{-1}$ ({\em bottom four panels}). 
In both cases the mean velocity $v$ ({\em upper left panel\/}), velocity dispersion $\sigma$ ({\em upper right panel\/}), and higher-order moments $h_3$ ({\em bottom left panel\/}) and $h_4$ ({\em bottom right panel\/}) of the LOSVD of two counter-rotating components are known as a function of their input velocity separation $\Delta V$. 
The points correspond to the measured values for mock spectra with $f_{\rm cr}=0.17$ ({\em red\/}), $0.23$ ({\em orange\/}), $0.33$ ({\em light-blue\/}), and $0.50$ ({\em black\/}), respectively. 
The points are connected to highlight the trend. 
The {\em black solid lines} indicate the input value of the mean velocity ($v=V_{\rm pro}$), velocity dispersion ($\sigma=\sigma_{\rm pro}$), and $h_3=h_4=0$ as a function of $\Delta V$. The {\em black dashed lines\/} indicate the default limits to the higher-order moments.}
\label{fig:mocktrend}
\end{figure*}
In Fig.~\ref{fig:mocktrend} we show the measured GH parameters $v$, $\sigma$, $h_3$, and $h_4$ as a function of the input velocity separation $\Delta V$, which are obtained from noise-free ({\em top panels\/}) and noisy ({\em bottom panels\/}) spectra with $\sigma_{\rm cr}/\sigma_{\rm pro}=1.0$. 
These trends show how the different fraction of the counter-rotating component causes: 1) a shift of $v$ from zero to the mean velocity of the primary (i.e., dominant) component $V_{\rm pro}$; 2) an increase in $\sigma$ starting from the input value $\sigma_{\rm pro}$; 3) discernible Gaussian deviations that are given by the non-null values of $h_3$ and $h_4$.
The most interesting feature is the sudden change that happens in all four parameters when $f_{\rm cr}\geq0.33$ and $\Delta V$ is large ($>150$ km s$^{-1}$). 
In particular, the parameter $v$ jumps from a low velocity to a higher value, closer to $V_{\rm pro}$, while the value of  $\sigma$ drops and then increases again.
As expected, $h_3$ anti-correlates with $v$, and $h_4$ correlates with $\sigma$. 
Both exceed values of $0.3$ that are usually not observed.
\citet{fabriciusKINEMATICSIGNATURESBULGES2012} studied the kinematics along the major axis of 45 S0-Sc galaxies and found 5 galaxies that exhibit large values of $h_3$ and $h_4$. 
The most extreme case is NGC~3521, which shows $|h_3|$ and $|h_4|$ as large as 0.24 and 0.35, respectively \citep[see also][]{zeilingerNGC3521Stellar2001}. 
For this reason it is thought to host a counter-rotating component; however,  by applying a spectral decomposition, \citet{coccatoSpectroscopicDecompositionNGC2018} found that the two distinct components are actually a bulge and disk in co-rotation.
They explained the counter-rotation as an artifact due to spurious signals that are generated in the Fourier space when the data have a poor spectral resolution. 
 \citet{zeilingerNGC3521Stellar2001} and \citet{fabriciusKINEMATICSIGNATURESBULGES2012} used the Fourier correlation
quotient method \citep{benderUnravelingKinematicsEarlytype1990,bender1994} for the LOSVD recovery of NGC~3521. 

We also found some ``off-trend points'' (e.g., the {orange} ones at large $\Delta V$) that are the results of a degenerate problem. 
Especially in the case where the two components have similar strength ({blue line}), the code returns different solutions depending on the initial guesses. 
For example, setting the initial guess of $v$ closer to the velocity of the counter-rotating component the measured $v$ is about zero. 
On the contrary, if a larger positive initial guess is used then $v$ is closer to the expected value following the trend.
In extreme cases, the solutions hit the positive boundary limit that is set to unity indicating that a better fit can be obtained by varying the initial guesses. 
For example, giving a $v$ that is closer to zero and/or a $\sigma$ that is larger than $\sigma_{\rm pro}$.
The jump from one solution to another is more evident for $f_{\rm cr}=0.50$ in the noisy spectra because here there is no dominant component that can bias the solution.
The noise also contributes to smoothing the best-fit solutions, as   can be seen in Fig.~\ref{fig:mocknoise}. 

For completeness, we analyze these trends when including the variation of $\sigma_{\rm cr}/\sigma_{\rm pro}$.
For a fixed value of $f_{\rm cr}$, we store the mock spectra mimicking a datacube, where the abscissa indicates the input velocity separation $\Delta V$, while the ordinate gives the input $\sigma_{\rm cr}/\sigma_{\rm pro}$,
but we do not find any specific trend.
However, the kinematic signatures of a counter-rotating component can be highlighted by extracting the stellar kinematics when adopting a single GH function: 1) measuring only $v$ and $\sigma$ makes clear the increase in the velocity dispersion, particularly evident for $f_{\rm cr}=0.50$; 2) including $h_3$ and $h_4$ gives a measurement of the strength of the counter-rotating component. The spectral fitting obtained with moments of order greater than four returns improved residuals, but it does not reflect any strong feature that could be helpful for the detection of counter-rotating components.
In particular, we  use the $h_3$ parameter to distinguish strong counter-rotation (corresponding to $|h_3|>0.2$) from weak counter-rotation ($|h_3| \leq 0.2$).

\section{Mock IFU observations}\label{sec:approach}
After having studied how the presence of two counter-rotating stellar components shapes the LOSVD for given values of $\Delta V$, $\sigma$, and $f_{\rm cr}$, it is necessary to investigate how this affects the kinematic map. 
In  a real galaxy we expect that $\Delta V$, $\sigma$, and $f_{\rm cr}$ vary with radius and position angle, producing typical patterns in the kinematic maps.
To this end, we simulate IFU observations of galaxies hosting two counter-rotating thin stellar disks. 

\subsection{Modeling counter-rotating galaxies}
To describe the physical properties of our galaxy models, we need to define their chemical composition (stellar population parameters), kinematics (LOS velocity and velocity dispersion), and photometry (surface brightness parameters). 

The galaxy chemical properties are given by the adopted stellar templates, which are characterized by age and metallicity.   
The template spectrum is chosen from a stellar library and convolved with a Gaussian LOSVD to match the kinematics along the LOS. 
During the convolution process also the instrumental characteristics (e.g., the instrumental line spread function) are taken into account.
At this early stage we decide to adopt the same stellar template for the two counter-rotating components, whereas the use of different stellar populations in our models and their effects on the kinematic signatures is left to a future analysis. 

The observed velocity field is generated projecting into the sky-plane an input rotation curve $V_{\rm rot}(R)$ typical of disk galaxies, which is characterized by a linear rotation in the central region and flat rotation at larger radii.
We define a circular disk of radius $R$ to be projected into the sky-plane considering the inclination $i$ and position angle $PA$ of the galaxy. 
The observed radius $r$ as function of the spatial coordinates ($x,y$) is expressed as 
\begin{equation}\label{eq:radius}
\begin{split}
r(x,y) = \Big \{ \Big [-(x-x_0)\sin{PA}+(y-y_0)\cos{PA} \Big ]^2 + \\ 
+ \Big [ \frac{-(x-x_0)\cos{PA}-(y-y_0)\sin{PA}}{\cos{i}} \Big ]^2 
\Big \}^{\frac{1}{2}}~,
\end{split}
\end{equation}
where ($x_0,y_0$) are the coordinates of the galaxy center.
The projected velocity field is given by
\begin{equation}
v_{\rm los}(x,y) = V_{\rm rot}(R) \cos{\theta} \sin{i}~,
\end{equation}
where
\begin{equation}
\cos{\theta} = \frac{-(x-x_0)\sin{PA}+(y-y_0)\cos{PA}}{r}~.
\end{equation}
To keep our models as simple as possible we assume $v_{\rm cr}(x,y)=-v_{\rm pro}(x,y)$ to describe the LOS velocity fields of the prograde and counter-rotating disks, while for their LOS velocity dispersion we assume $\sigma_{\rm cr}(x,y)=\sigma_{\rm pro}(x,y)=$ constant.

The spectra are re-scaled according to the given surface brightness profile $I(x,y)$. 
To describe the surface brightness of a disk component we adopt the exponential law \citep{freemanDisksSpiralGalaxies1970}
\begin{equation}
I(x,y) = I_0 \exp{\left(-\frac{r}{h} \right)}~,
\end{equation}
where $r$ is defined as in Eq.~\ref{eq:radius}, while $I_0$ and $h$ are the central surface brightness and scale length of the disk, respectively.
Disk isophotes are ellipses centered on the galaxy center ($x_0,y_0$) with constant $PA$ and constant ellipticity $\epsilon = 1-q^\prime$, where $q^\prime$ is the apparent axial ratio and it is related to galaxy inclination through the relation $q^\prime=\cos{i}$. 
The counter-rotating model will have $I_{\rm m}(x,y) = I_{\rm pro}(x,y)+I_{\rm cr}(x,y)$. 
The final datacube (i.e., galaxy model) is obtained by adding the contribution of the counter-rotating disk to that of the prograde disk, which is kept fixed. 
We ignore the contribution of a bulge component, which is very small in most of counter-rotating galaxies, and blurry due to the seeing, since our analysis focuses on the effects in outer region of the galaxy.
Furthermore, we do not add sky, detector, or Poisson noise, but they can be modeled and included a posteriori if required for the analysis.

\subsection{Validation of the counter-rotating galaxy models}\label{sec:test}
\begin{figure*}
\centering
\includegraphics[width=0.7\textwidth]{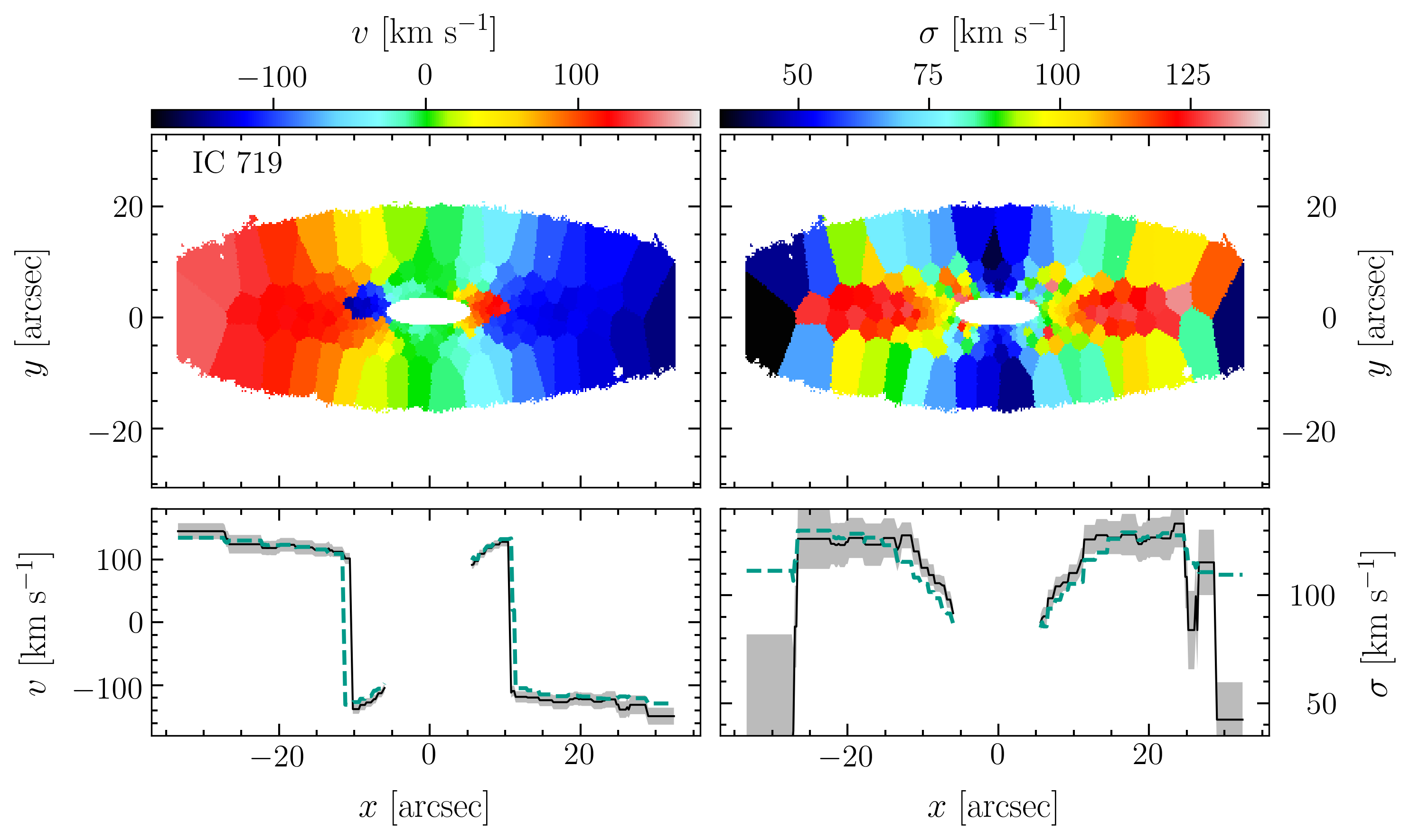}
\includegraphics[width=0.7\textwidth]{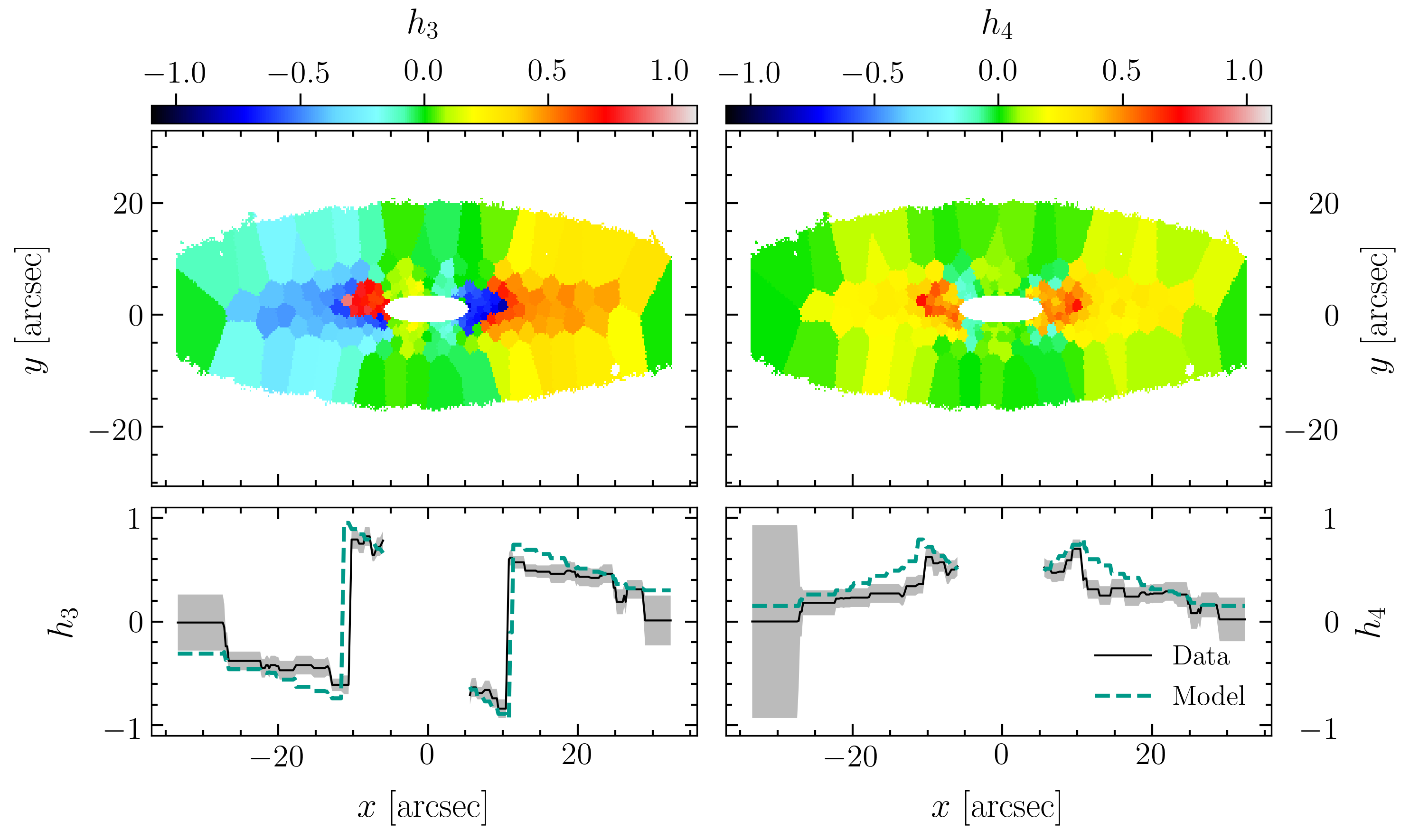}
\caption{Stellar kinematic maps ({\em top panels\/}) and major-axis radial profiles ({\em solid lines\/} with {\em gray error bars,}  {\em bottom panels\/}) of the mean velocity $v$ ({\em upper left panels\/}), velocity dispersion $\sigma$ ({\em upper right panels\/}), and higher-order moments $h_3$ ({\em lower left panels\/}) and $h_4$ ({\em lower right panels\/}) of the LOSVD for IC~719 derived from MUSE data. The {\em green dashed lines\/} show the major-axis radial profiles of the kinematic parameters obtained with our model.}
\label{fig:IC719}
\end{figure*}

To test the reliability of our approach, we built a kinematic model of IC~719 for which the photometric, kinematic, and stellar population parameters needed to construct the mock data are available in \citet{pizzellaEvidenceFormationYoung2018}. 

The decomposition of IC~719 provides very detailed information on the two counter-rotating components that allows us to drop some assumptions previously made to generate our model. For example, we do not use the same stellar population model and equal kinematics for the two counter-rotating components. 
We use instead two different E-MILES SSP models, one for the main (prograde) disk with mean ages $t_{\rm pro}=3.5$ Gyr and $t_{\rm cr}=1.5$ Gyr for the secondary (counter-rotating) one. 
We choose mean metallicities $[\rm{M/H}]_{\rm pro}=+0.26$ dex and $[\rm{M/H}]_{\rm cr}=-0.25$ dex, consistent with the available values for the SSP models and measured values at 1.7 kpc. 
The velocity fields are generated adopting two rotation curves that best reproduce those extracted along the major axis of IC~719, while the velocity dispersion fields are assumed to be constant with values of \mbox{60 km s$^{-1}$} and \mbox{40 km s$^{-1}$} for $\sigma_{\rm pro}$ and $\sigma_{\rm cr}$, respectively.  
The surface brightness distribution of both disks follows an exponential law, although with different scale length $h_{\rm pro}=1.5$ kpc and $h_{\rm cr}=1.0$ kpc. 
The counter-rotating components equally contribute to the galaxy surface brightness at about 1.0 kpc.  
We use the relative flux contribution to constrain the central surface brightness of the main component $I_{\rm 0,pro}$, which is expressed in arbitrary linear units, while the central contribution of the counter-rotating disk is given as a fraction of the main one (i.e., $I_{\rm 0,cr}=f_{0} \times I_{\rm 0,pro}$).
In the innermost region ($<0.8$ kpc) the parameters are poorly constrained, thus we do not consider the central region and we do not even include the bulge contribution. 
Indeed, IC 719 has a very small bulge.

To directly compare data and model stellar kinematics we spatially re-sample the MUSE datacube of IC~719 with the same binning map. 
We use the map obtained by the Voronoi tesselletion method developed by \citet{cappellariAdaptiveSpatialBinning2003}. 
We choose a target signal-to-noise $(S/N)_{\rm T}=90$ and select only spaxels with $S/N>7$. 
This allows us to avoid poor quality spaxels, which could introduce undesired systematic effects in the data at low surface-brightness regimes that are not straightforward to account for.
We then extract the stellar kinematics with pPXF limiting the wavelength range to 8400--8800 \AA. 
This short wavelength range permits us to decrease the computational time for the measurements and also allows us to assume that the instrumental line spread function does not vary as function of wavelength.
We perform the kinematic measurements following the same scheme used in Sect.~\ref{sec:LOSVD}. The uncertainties are the 1$\sigma$ formal errors returned by pPXF.
The first measurements, obtained fitting only $v$ and $\sigma$, were subsequently used as starting guesses for $v$ and $\sigma$, when we also fit the higher-order moments $h_3$ and $h_4$ of the LOSVD. 
The best-fit solution of each bin is visually inspected to check the goodness of the fit and when a better solution is expected we re-perform the measurements varying the initial guesses.

We show in Fig.~\ref{fig:IC719} the stellar kinematic maps of IC~719 and the good agreement between the radial profiles of the observed and model kinematics extracted along the galaxy major axis within 1 arcsec aperture. 
In particular, we nicely recover the reversal of $v$ measured at $|r|\sim10$ arcsec from the center, the remarkable $2\sigma$ feature with $\sigma$ peaking at $\sim130$ km~s$^{-1}$ for $10<|r|<25$ arcsec, and the strongly asymmetric LOSVD with the higher-order moments steadily rising to $|h_3|\sim1.0$ and $h_4\sim0.8$ at $|r|\sim10$ arcsec. 
We note that we do not fit the MUSE data, but plot the stellar kinematics resulting from the kinematic model built from the known parameters of IC~719.

\subsection{Mock counter-rotating galaxies}\label{sec:data}

We built our mock counter-rotating galaxies assuming the same distance for all models to make possible their direct comparison. 
We fix the spatial sampling $\Delta x = 50$ pc pixel$^{-1}$, which corresponds to a distance of 51.6 Mpc considering the MUSE spatial sampling of 0.2 arcsec pixel$^{-1}$.
We decided to have the $PA=90^\circ$ such that the major axis of the galaxy model is aligned with the direction of the $x$-axis.
For these choices, a disk scale length of 1.0 kpc is  located at the radius of 20 pixels along the $x$-axis. 
Within the field of view (FoV) of 60 arcsec, we cover a physical dimension of 15 kpc.

We model the two counter-rotating components with the same SSP model of age $t=6$ Gyr and metallicity ${\rm[M/H]}=-0.25$ dex.
We consider a rotation curve with a maximum velocity of 
100 km s$^{-1}$.
Adopting three different inclinations $i=40^\circ,~ 50^\circ,$ and $70^\circ$ for the two counter-rotating stellar disks their projected maximum velocity separations are $\sim130$ km s$^{-1}$, $\sim150$ km s$^{-1}$, and $\sim190$ km s$^{-1}$, respectively. 
These $\Delta v$ address the cases in which the spectral lines are not strongly double-peaked.
For both stellar disks we opt for a constant velocity dispersion $\sigma_{\rm pro} = \sigma_{\rm cr} = 55$ km s$^{-1}$ all over the FoV. 
This value is consistent with that observed at 10 arcsec for the main component of IC~719.

For the photometric parameters of the galaxy models 
we fix the central surface brightness $I_{0,\rm pro}$ and scale length $h_{\rm pro}=1.5$ kpc of the prograde disk. 
We adopt eight different values of the multiplicative factor $f_{\rm 0}$ such that the central surface brightness of the counter-rotating component can be expressed relative to that of the prograde component, as in Sect.~\ref{sec:test}.
The ratio of the surface brightness of the counter-rotating disk to the surface brightness of the model (i.e., $f_{\rm cr}(x,y)=I_{\rm cr}(x,y)/I_{\rm m}(x,y)$) spatially varies over the FoV, ranging from 0 to 1.
We also consider three different values for the scale length of the counter-rotating disk $h_{\rm cr}=1, 1.5,$ and 2 kpc.
The ratio of the total luminosity of the counter-rotating disk to that of the galaxy model, $L_{\rm cr}/L_{\rm T}$, varies between 0.10 and 0.78 depending on $h_{\rm cr}$.

We divide our models into three sets of 24 galaxies each depending on their inclination. Each set of galaxy models is then divided into three subsets according to the scale lengths of the two counter-rotating disks: subset 1 includes the models with more concentrated counter-rotating disks dominating the surface brightness distribution of the galaxy in the inner regions ($h_{\rm cr}<h_{\rm pro}$); subset 2 has the models with constant ratio of the surface brightness of the counter-rotating disk to that of the prograde disk at all radii ($h_{\rm cr}=h_{\rm pro}$); subset 3 counts models with a counter-rotating disk dominating the outer surface brightness distribution of the galaxy ($h_{\rm cr}>h_{\rm pro}$).
The photometric parameters characterizing the galaxy models are given in Table~\ref{tab:parmod}. The naming convention ({\tt miinnn}) provides the inclination of the galaxy in degrees ({\tt ii}) and a sequential number ({\tt nnn}) to identify the models. For instance, {\tt m40001} is the first model with inclination $i=40^{\circ}$.

\begin{table*}
\centering  
\caption{Parameters of the galaxy models.}      \label{tab:parmod}      
{\relsize{0}   
\begin{tabular}{c c c c c c c c }
\hline \hline
\rule{0pt}{4ex}
Subset & Model & \Large{$\frac{I_{\rm 0, cr}}{I_{\rm 0, pro}}$} & \Large{$\frac{I_{\rm 0, cr}}{I_{\rm 0, m}}$} & \Large{$\frac{h_{\rm cr}}{h_{\rm pro}}$} &  \Large{$\frac{ L_{\rm cr} }{ L_{\rm pro} }$} &  \Large{$\frac{ L_{\rm cr} }{ L_{\rm T}}$} & 
{$r_{\rm e}$}  \\
\rule[-1ex]{0pt}{0pt} 
&&&&&&& [arcsec] \\
(1) & (2) & (3) & (4) & (5) & (6) & (7) & (8) \rule[-1ex]{0pt}{0pt}\\
\hline
\rule{0pt}{2.5ex}
\multirow{ 8}{*}{\#1}  &{\tt mii001}  & 0.25  & 0.20 & 0.67 & 0.11 & 0.10 &  9.6 \\
                       &{\tt mii002}  & 0.50  & 0.33 & 0.67 & 0.22 & 0.18 &  9.3 \\ 
                       &{\tt mii003}  & 0.75  & 0.43 & 0.67 & 0.33 & 0.25 &  9.0 \\ 
                       &{\tt mii004}  & 1.00  & 0.50 & 0.67 & 0.44 & 0.31 &  8.8 \\ 
                       &{\tt mii005}  & 1.25  & 0.56 & 0.67 & 0.54 & 0.35 &  8.6 \\ 
                       &{\tt mii006}  & 1.50  & 0.60 & 0.67 & 0.65 & 0.39 &  8.5 \\ 
                       &{\tt mii007}  & 1.75  & 0.64 & 0.67 & 0.76 & 0.43 &  8.4 \\ 
                       &{\tt mii008}  & 2.00  & 0.67 & 0.67 & 0.87 & 0.47 &  8.2 \\
                    \hline                          
                    \rule{0pt}{2.5ex}               
\multirow{ 8}{*}{\#2}  &{\tt mii009}  & 0.25  & 0.20 & 1.00 & 0.25 & 0.20 & 10.0 \\ 
                       &{\tt mii010}  & 0.50  & 0.33 & 1.00 & 0.50 & 0.33 & 10.0 \\ 
                       &{\tt mii011}  & 0.75  & 0.43 & 1.00 & 0.75 & 0.43 & 10.0 \\ 
                       &{\tt mii012}  & 1.00  & 0.50 & 1.00 & 1.00 & 0.50 & 10.0 \\ 
                       &{\tt mii013}  & 1.25  & 0.56 & 1.00 & 1.25 & 0.56 & 10.0 \\ 
                       &{\tt mii014}  & 1.50  & 0.60 & 1.00 & 1.50 & 0.60 & 10.0 \\ 
                       &{\tt mii015}  & 1.75  & 0.64 & 1.00 & 1.75 & 0.64 & 10.0 \\ 
                       &{\tt mii016}  & 2.00  & 0.67 & 1.00 & 2.00 & 0.67 & 10.0 \\
                    \hline                          
                    \rule{0pt}{2.5ex}               
\multirow{ 8}{*}{\#3}  &{\tt mii017}  & 0.25  & 0.20 & 1.33 & 0.44 & 0.31 & 11.0 \\ 
                       &{\tt mii018}  & 0.50  & 0.33 & 1.33 & 0.88 & 0.47 & 11.5 \\ 
                       &{\tt mii019}  & 0.75  & 0.43 & 1.33 & 1.33 & 0.57 & 11.8 \\ 
                       &{\tt mii020}  & 1.00  & 0.50 & 1.33 & 1.77 & 0.64 & 12.0 \\ 
                       &{\tt mii021}  & 1.25  & 0.56 & 1.33 & 2.21 & 0.69 & 12.2 \\ 
                       &{\tt mii022}  & 1.50  & 0.60 & 1.33 & 2.65 & 0.73 & 12.4 \\ 
                       &{\tt mii023}  & 1.75  & 0.64 & 1.33 & 3.10 & 0.76 & 12.5 \\ 
                       &{\tt mii024}  & 2.00  & 0.67 & 1.33 & 3.54 & 0.78 & 12.6 \\ 
\hline  \rule{0pt}{1ex}
\end{tabular}\\
\begin{minipage}{\textwidth}\small
{\bf Notes.} Col.(1): Subset number. Col.(2): Name of the galaxy model, where ${\tt ii}=\{40, 50, 70\}$ is the model inclination in degree. Col.(3-4): Ratio of the central surface brightness of the counter-rotating disk to that of the prograde disk and to that of the model. Col.(5): Ratio of the scale length of the counter-rotating disk to that of the prograde disk. Col.(6-7): Ratio of the total luminosity of the counter-rotating disk to that of the prograde disk and to that of the model. Col.(8): Effective radius of the model.
\end{minipage}  
}
\end{table*}

\section{Analysis of stellar kinematic maps}\label{analysis}

For all our models we perform a standard procedure that can be automatically run for the entire set.
In this analysis we apply a spatial re-sampling of the datacubes. and then from each bin we extract the stellar kinematics, as done in Sect.~\ref{sec:test}.

Since our models are noise-free, in principle, our data already have enough $S/N$ to perform the spectral fit, but we want to spatially re-sample the datacube as is  usually done for real observations.
For this reason, when applying the Voronoi binning, we assume that the noise associated with our data is Poissonian ($N\approx\sqrt{S}$).
With this assumption it turns out that for all models $(S/N)_{\rm T}=150$ is a good compromise between our ability to measure the kinematics and the dimension and spatial distribution of the bins. 
This also avoids effects due to the averaging of spectra over a wide galaxy region. 
After extracting the stellar kinematics we first look at the qualitative effects due to the presence of the counter-rotating component on the kinematic maps, and then we perform a more quantitative analysis that  consists in an automatic detection of the observed feature along the major axis radial profiles of all models.
We present   the two-moment analysis (Sect.~\ref{sec:2mom}) and the four-moment analysis (Sect.~\ref{sec:4mom}).

\subsection{Counter-rotating signatures from two-moment analysis}\label{sec:2mom}

\begin{figure*}
    \centering
    \includegraphics[width=1\textwidth]{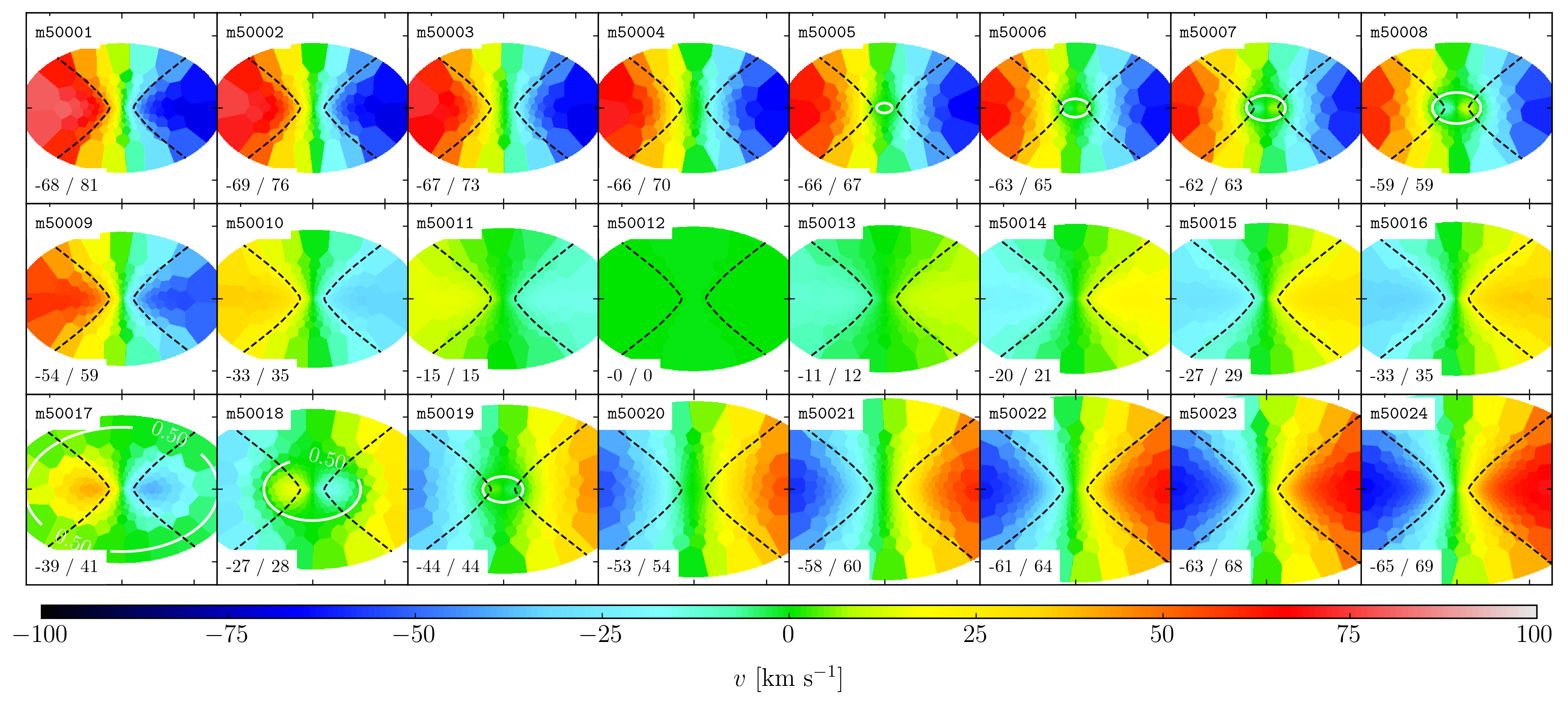}\\
    \includegraphics[width=1\textwidth]{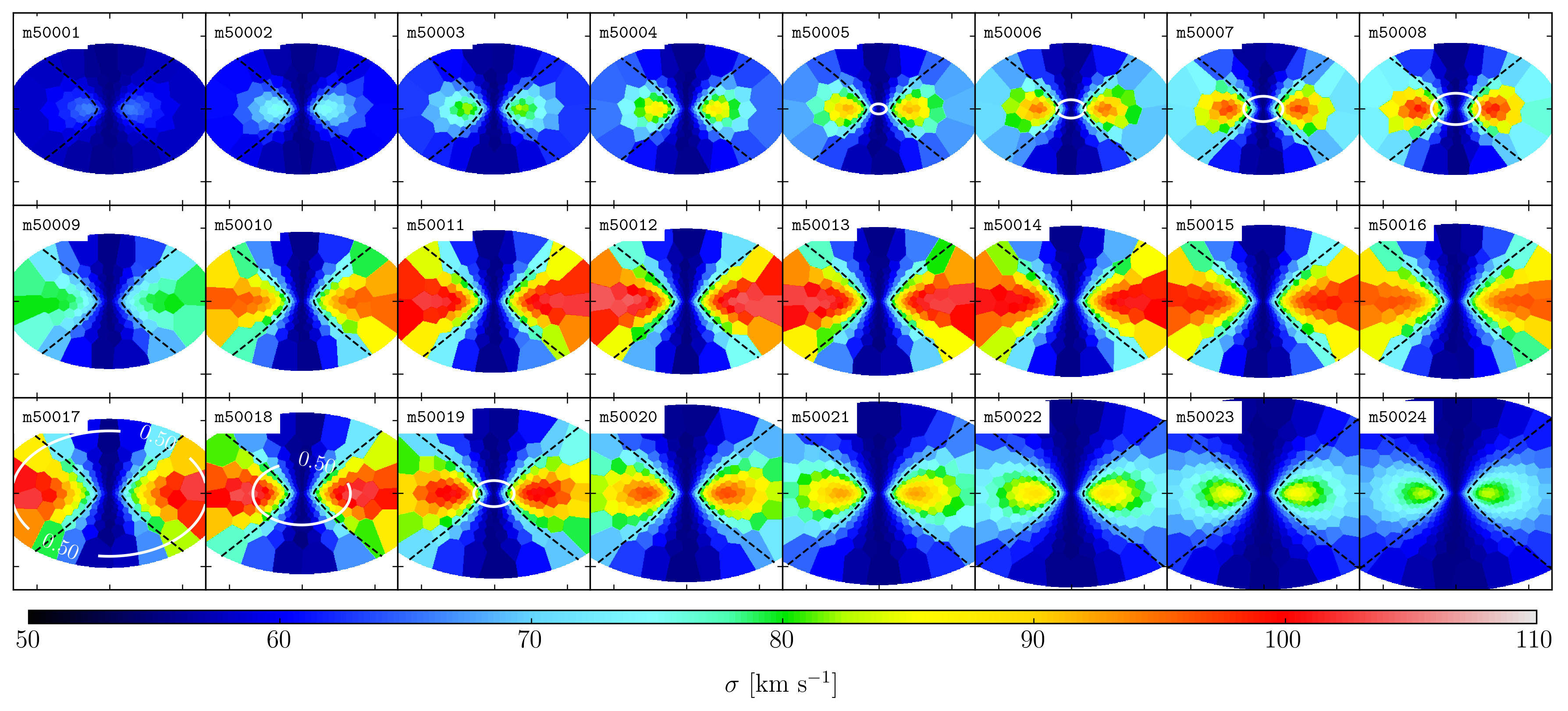}\\
    \caption{Stellar velocity ({\em top panels\/}) and velocity dispersion ({\em bottom panels\/}) maps of models belonging to subset 1 ({\em top row}), 2 ({\em middle row}), and 3 ({\em bottom row}) with an inclination of~ 50$^\circ$. 
    The {\em white ellipse} shows where $f_{\rm cr}=0.50$; this level is not reached in all models.
    The {\em black dashed lines} indicate the velocity separation $\Delta v$ = 100 km s$^{-1}$. 
    In the {\em bottom left corner} of each velocity map we give the minimum and maximum values of $v$.
    These kinematic measurements are obtained fitting only $v$ and $\sigma$. 
    }
    \label{fig:2vmod50}
\end{figure*}

The kinematic maps obtained fitting only $v$ and $\sigma$ are shown in Fig.~\ref{fig:2vmod50}. 
By inspecting these maps we confirm that the decrease in the mean velocity and double-peaked velocity dispersion are the main kinematic features due to the presence of two counter-rotating stellar disks. 

The decrease in the measured parameter $v$ is observed because the mean velocity of the galaxy model is always biased towards the velocity of the dominant component being the one that contributes more than half of the total surface brightness.
Thus, the slope and amplitude of the rotation curve decrease with the increasing contribution of the counter-rotating disk.
The inversion of the velocity field occurs at the radius corresponding to $f_{\rm cr}=0.50$ (see, e.g., {\tt m50008}).  
In subset 2, $f_{\rm cr}$ is constant over the entire FoV, thus we do not observe a change of sign within the same velocity map, but the direction of rotation changes for models {\tt m50013-16} compared to the velocity field of models {\tt m50009-11}.
For   model {\tt m50012} we find a constant null velocity field because the prograde and counter-rotating disks have equal surface brightness at all radii and give the same contribution to the model ($h_{\rm cr}=h_{\rm pro}$ and $f_{\rm cr}=0.50$). 
In subset 3, for models with $f_{\rm 0}<1.00,$ the counter-rotating disk dominates the outer regions, while in the other cases the counter-rotating disk becomes the dominant component at all radii ({\tt m50020-24}). 
This last subset can be considered  the reverse situation of subset 1 with effects on larger scales.

The size, shape, and slope of the two peaks seen in the velocity dispersion map strongly depend on the velocity separation and contribution of the two counter-rotating stellar disks. 
It is worth clarifying that we consider as  the $2\sigma$ feature also what we observe in the models of subset 2, although there is no decrease in $\sigma$ after the maxima. 
This is because 1) both stellar disks equally contribute to the model at all radii and 2) at large distances from the center the rotation curves become flat (i.e., the $\Delta v$ remains constant). 
Therefore, we observe a biconical shape instead of two symmetric, off-center, roundish regions of higher velocity dispersion.
We consider the $2\sigma$ feature as the most important kinematic diagnostic for counter-rotating galaxies because it appears even for galaxy models with a regular rotation and no evidence of velocity lowering or reversal. 
This is the case of {\tt m50004}, as  can be seen from its velocity and velocity dispersion maps in Fig.~\ref{fig:2vmod50}. 
In subset 3 the $2\sigma$ peaks are more elongated than those in subset 1 because the counter-rotating contribution is higher in the outermost than in the innermost regions.

\begin{figure*}
\centering
\centering
\includegraphics[width=0.33\textwidth]{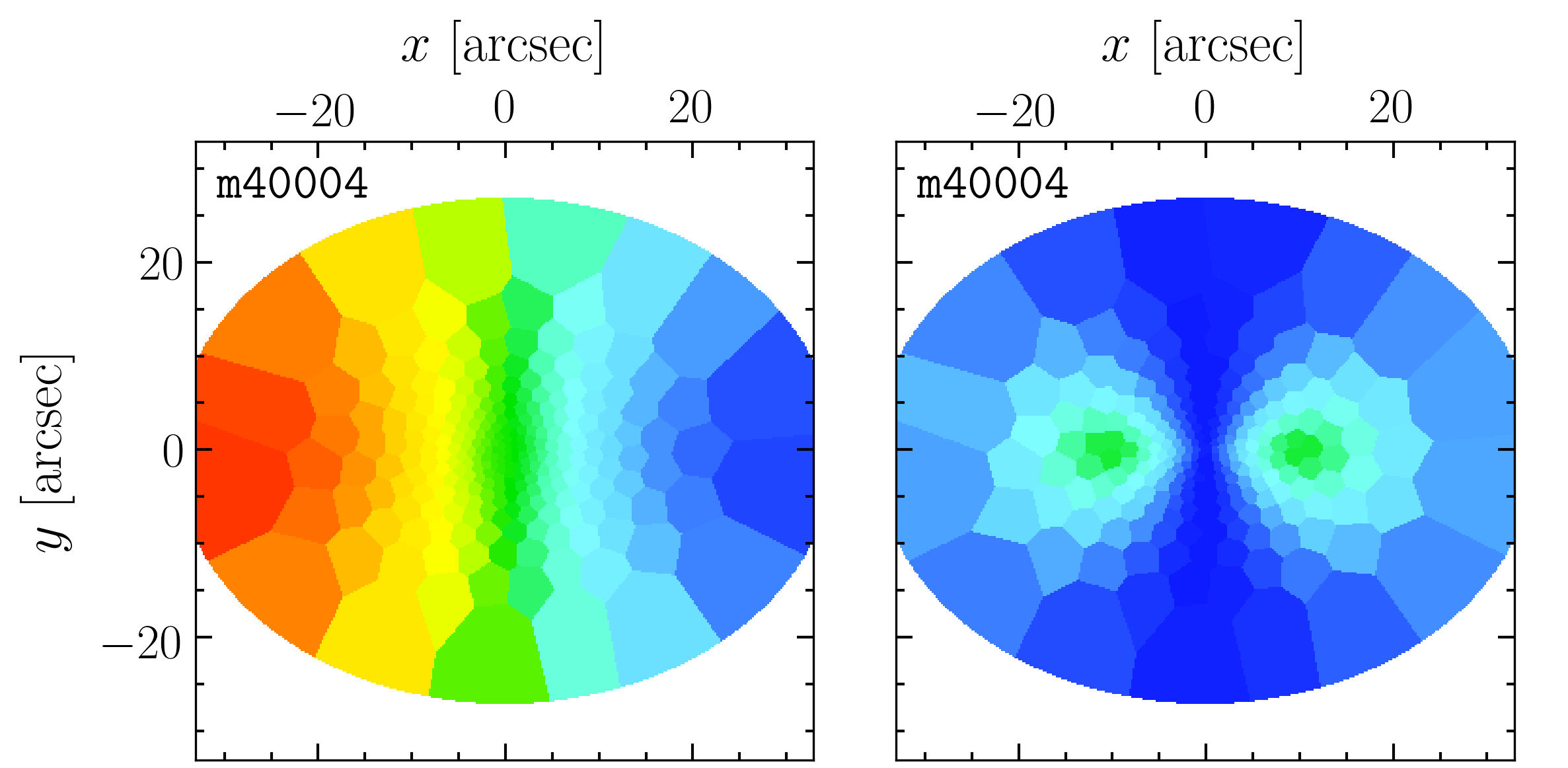}
\includegraphics[width=0.33\textwidth]{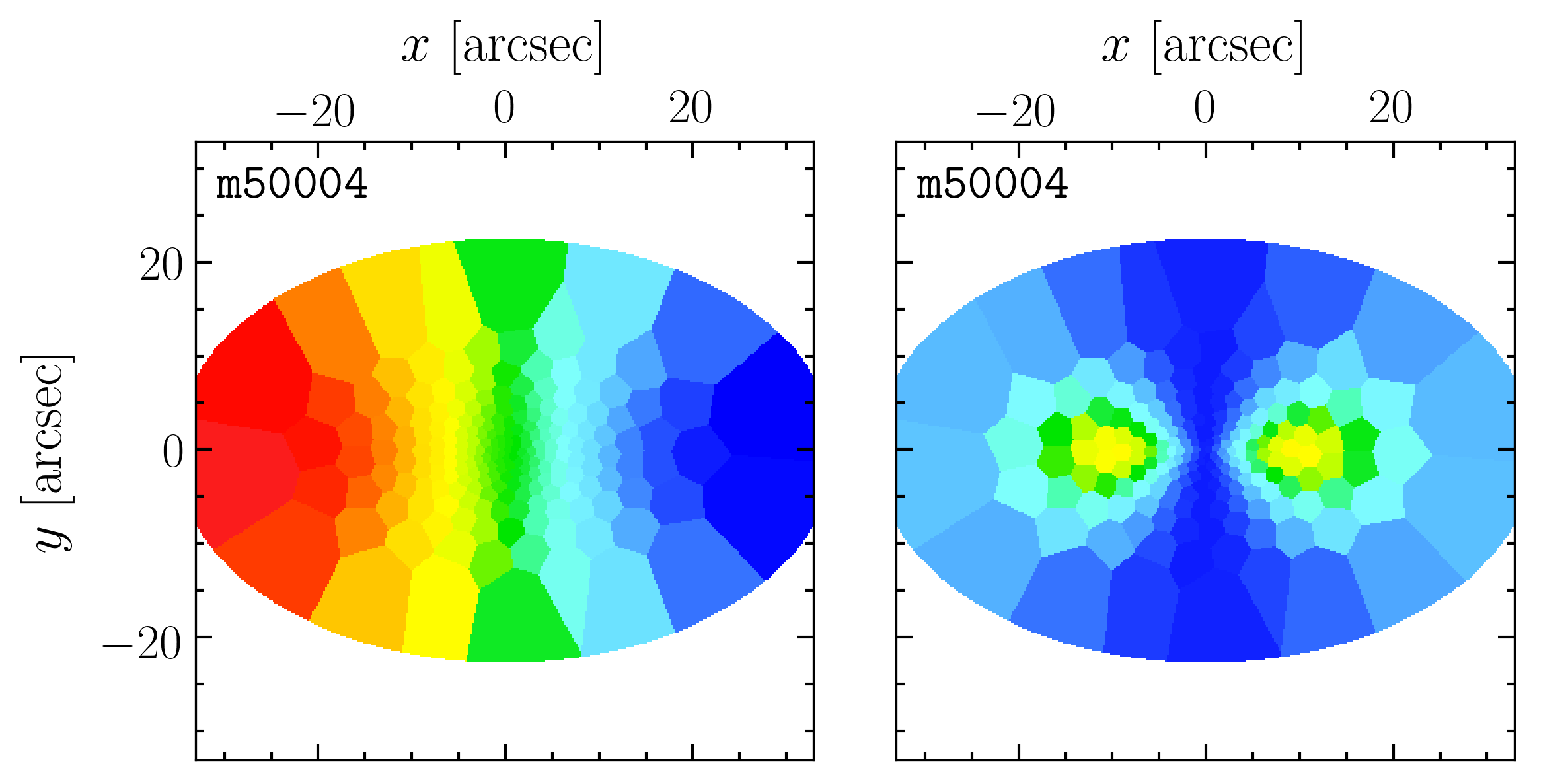}
\includegraphics[width=0.33\textwidth]{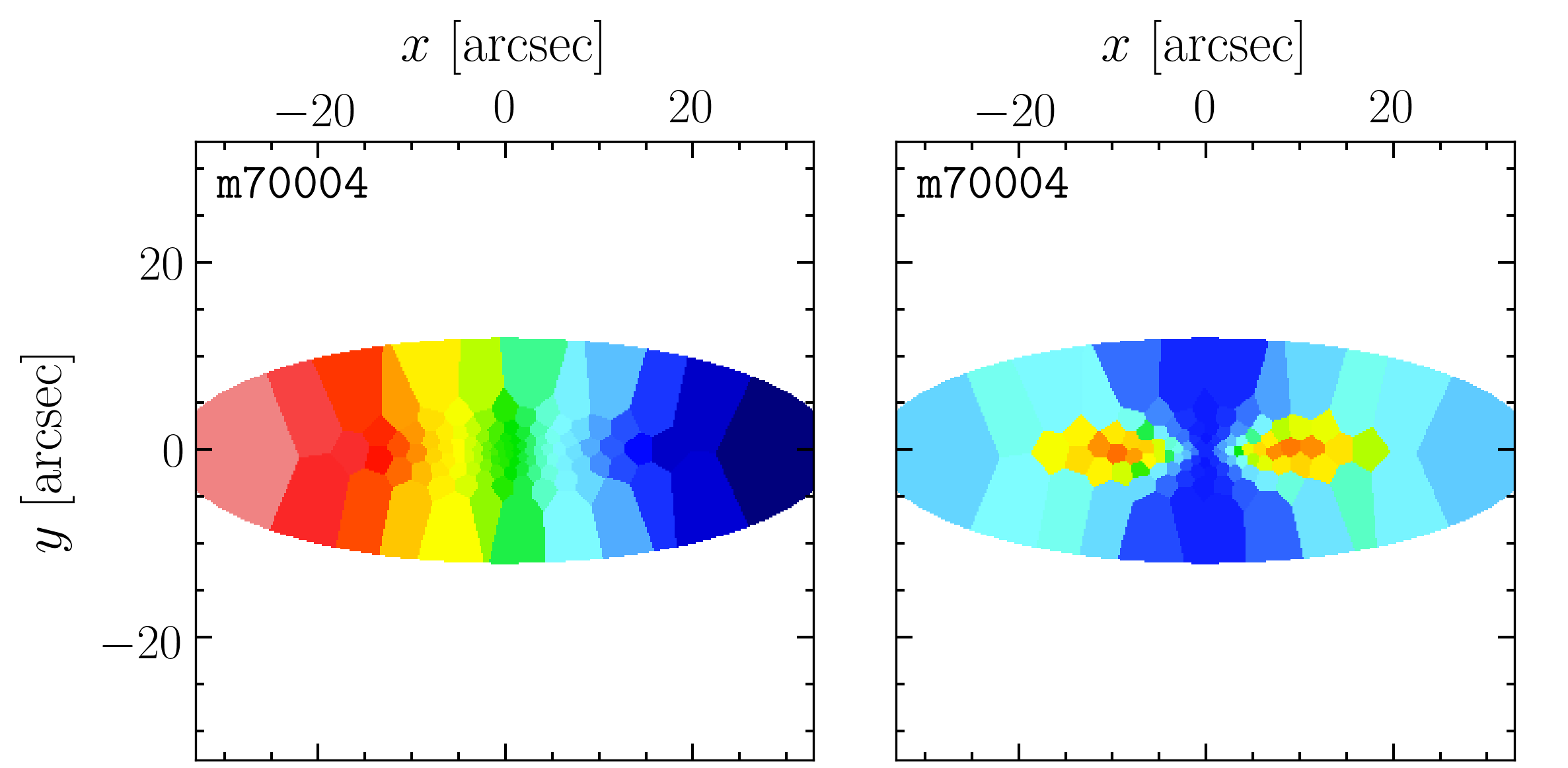}\\
\centering
\includegraphics[width=0.33\textwidth]{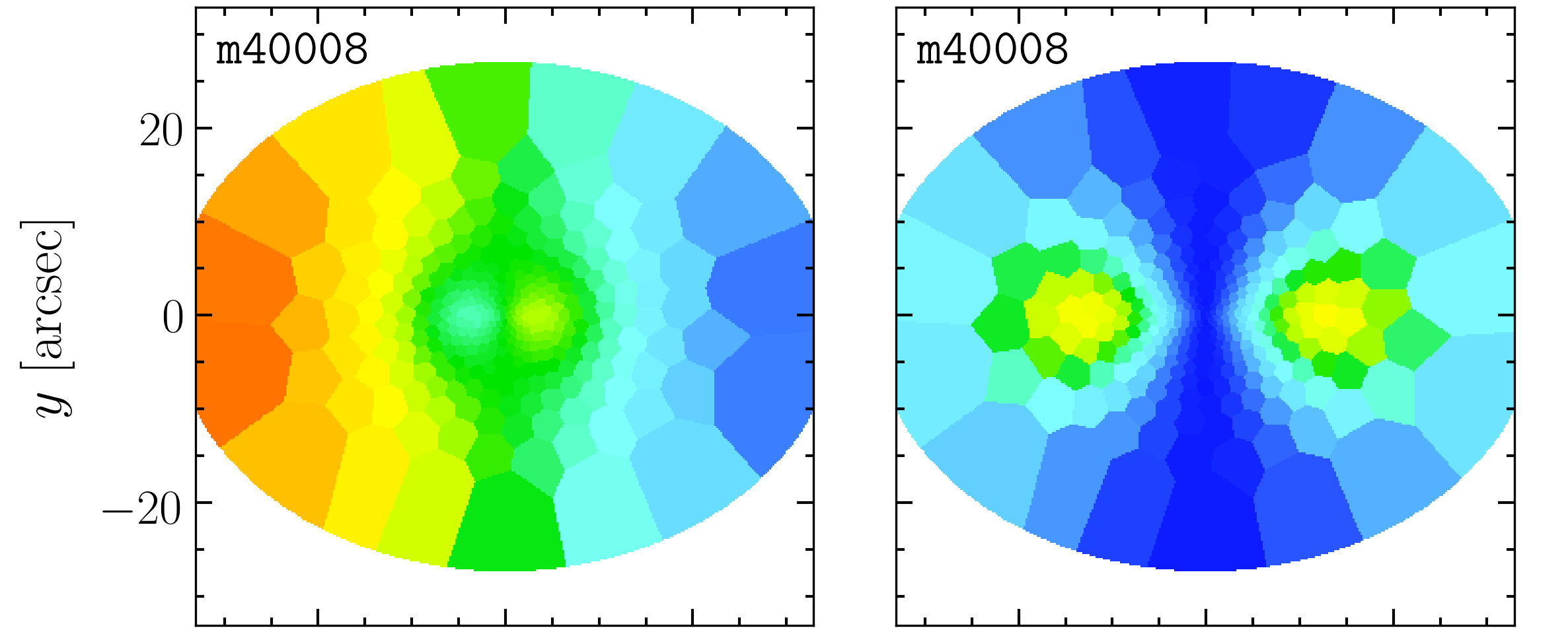}
\includegraphics[width=0.33\textwidth]{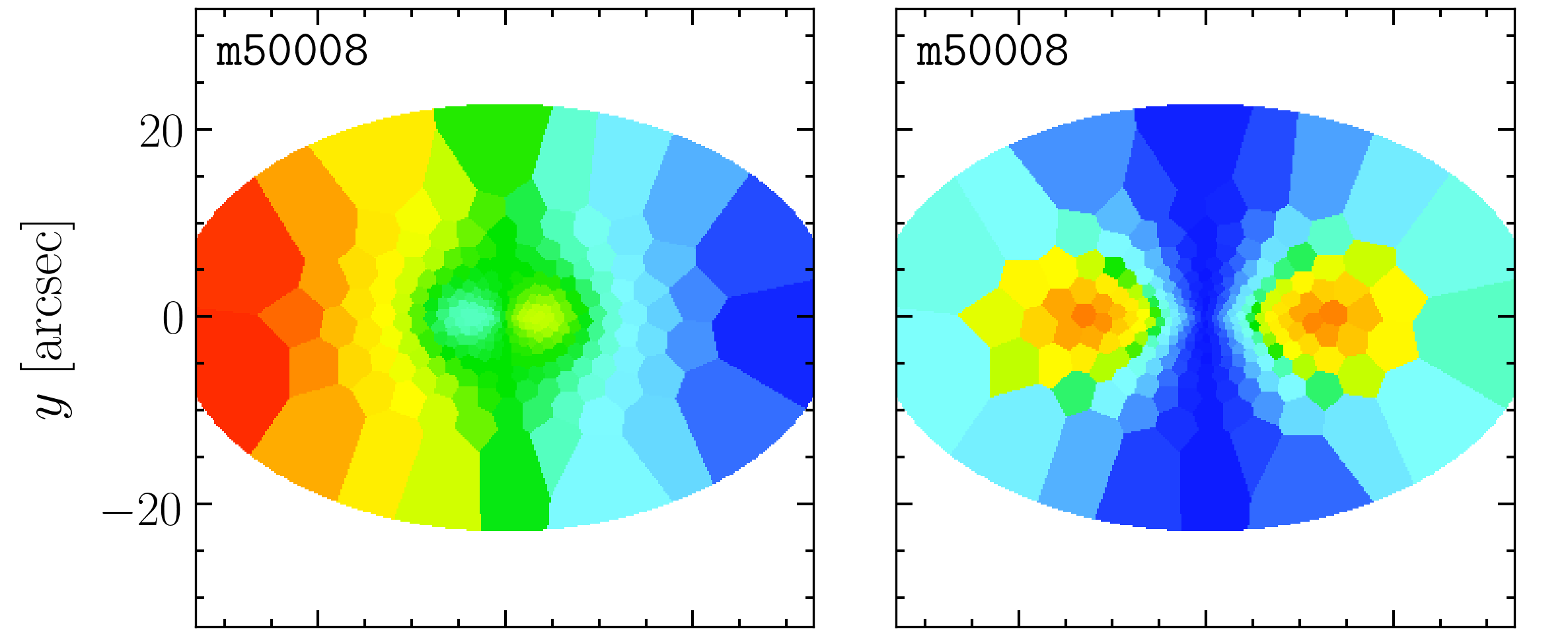}
\includegraphics[width=0.33\textwidth]{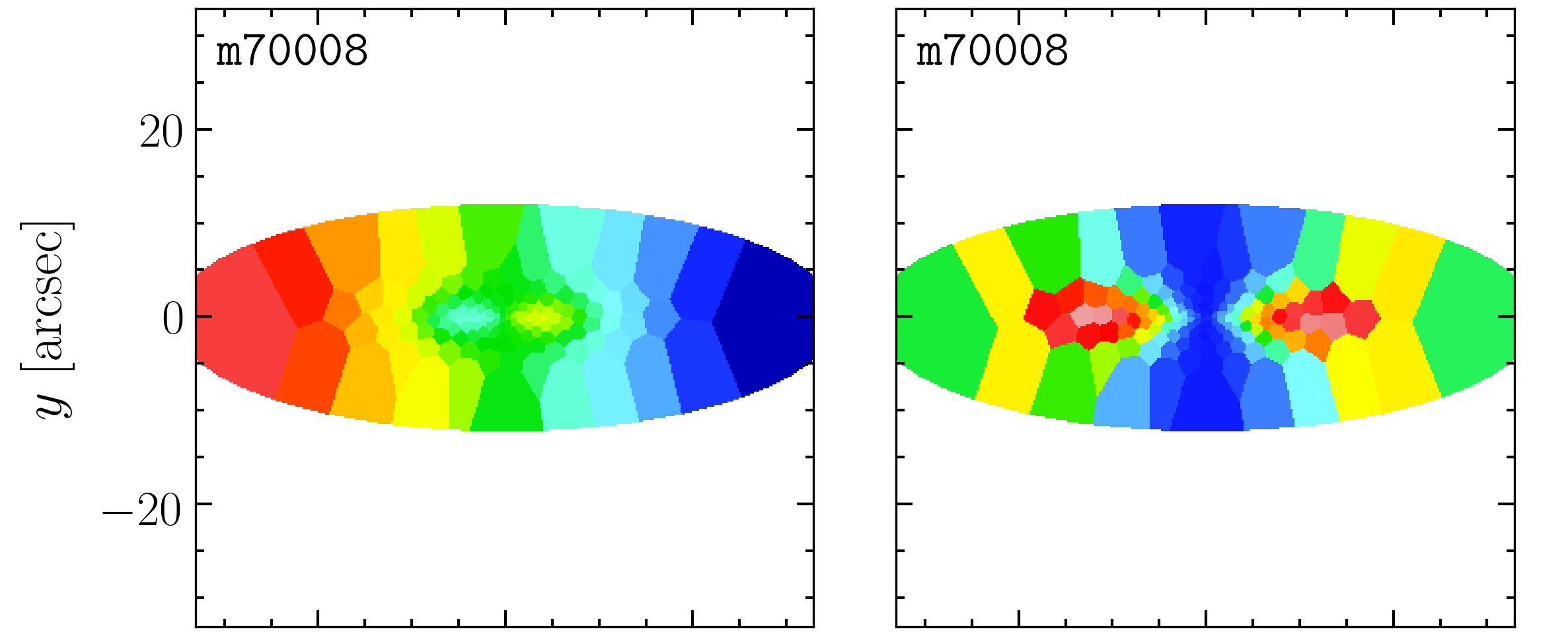}\\
\centering
\includegraphics[width=0.33\textwidth]{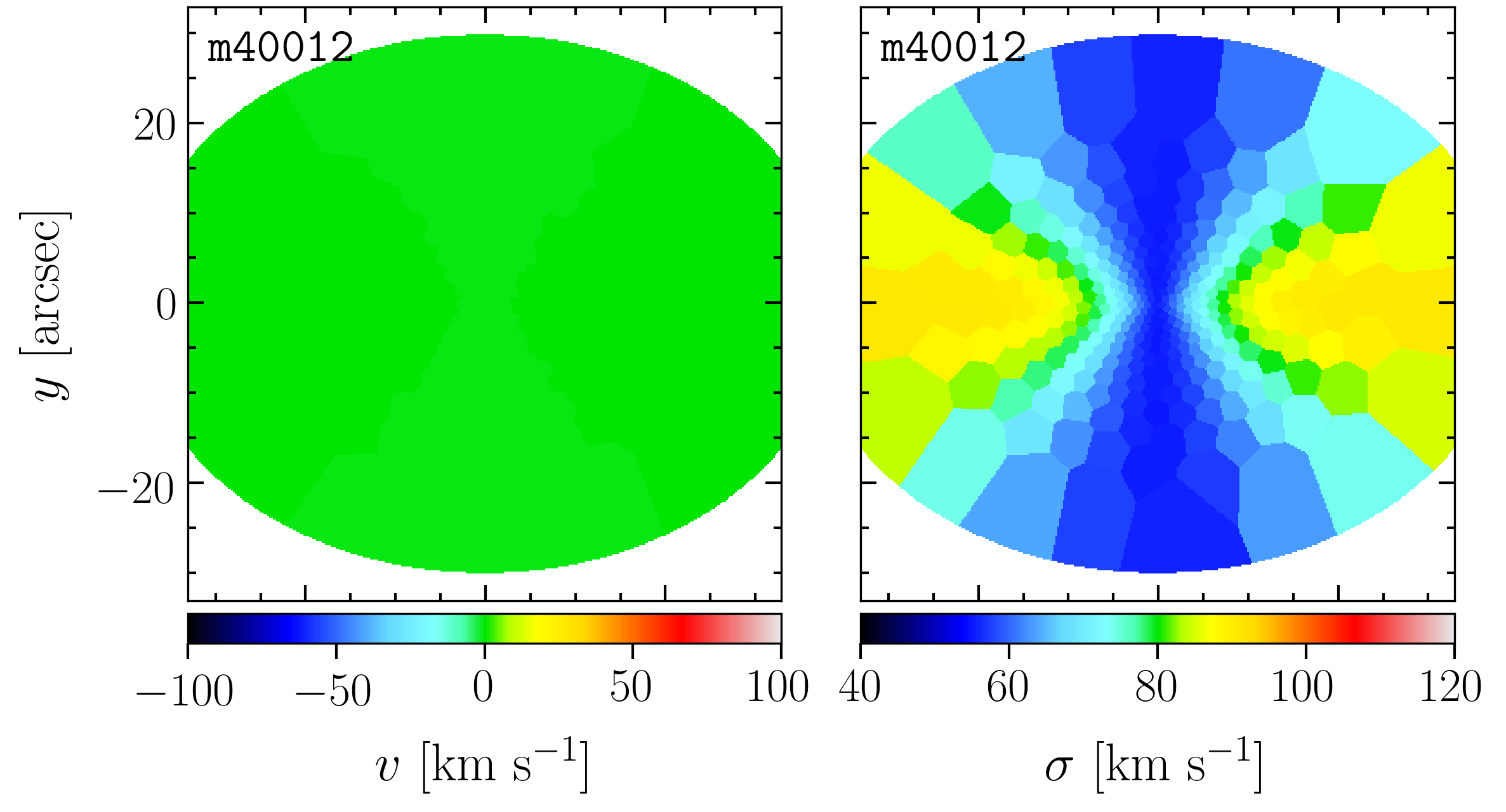}
\includegraphics[width=0.33\textwidth]{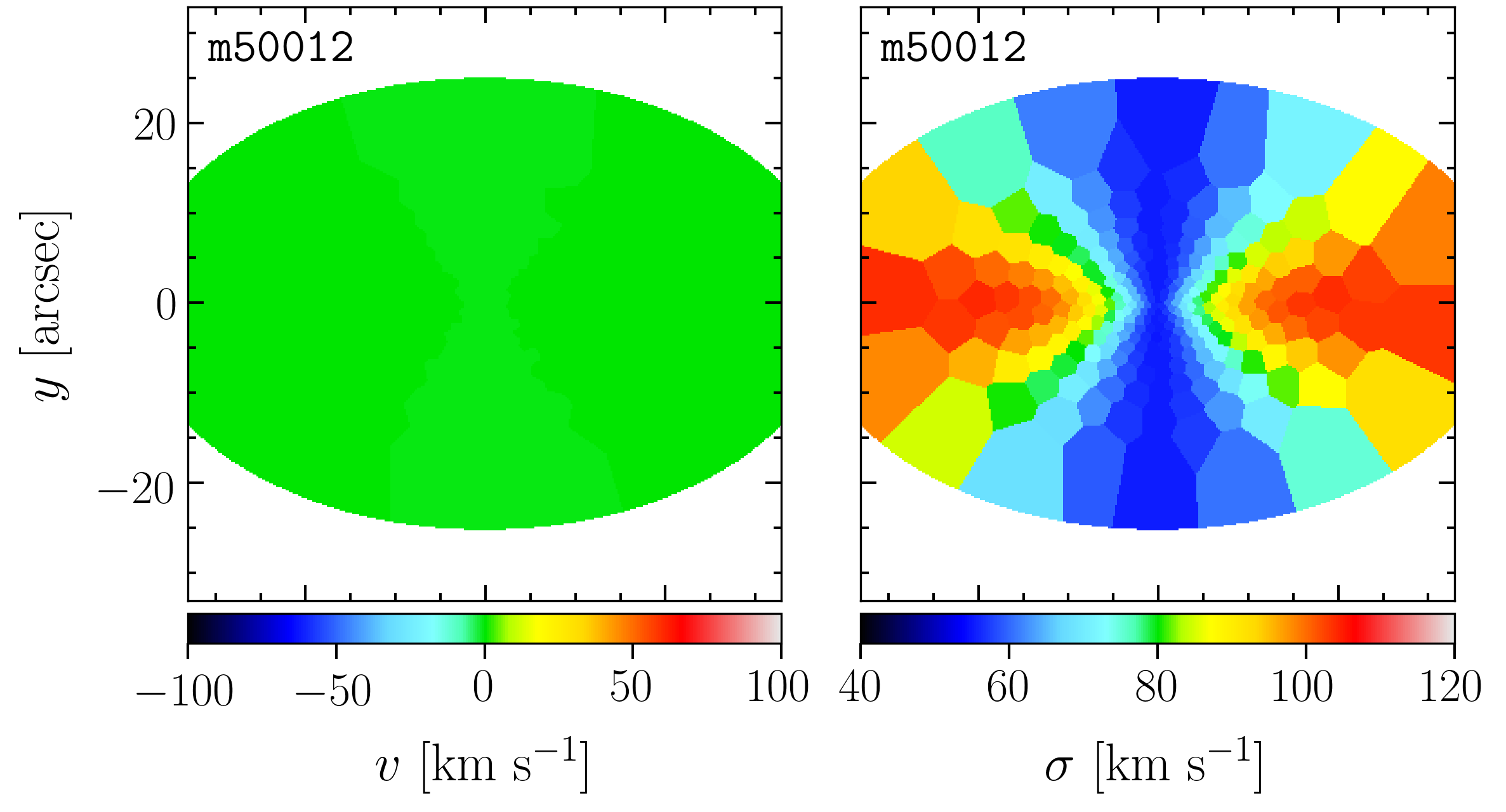}
\includegraphics[width=0.33\textwidth]{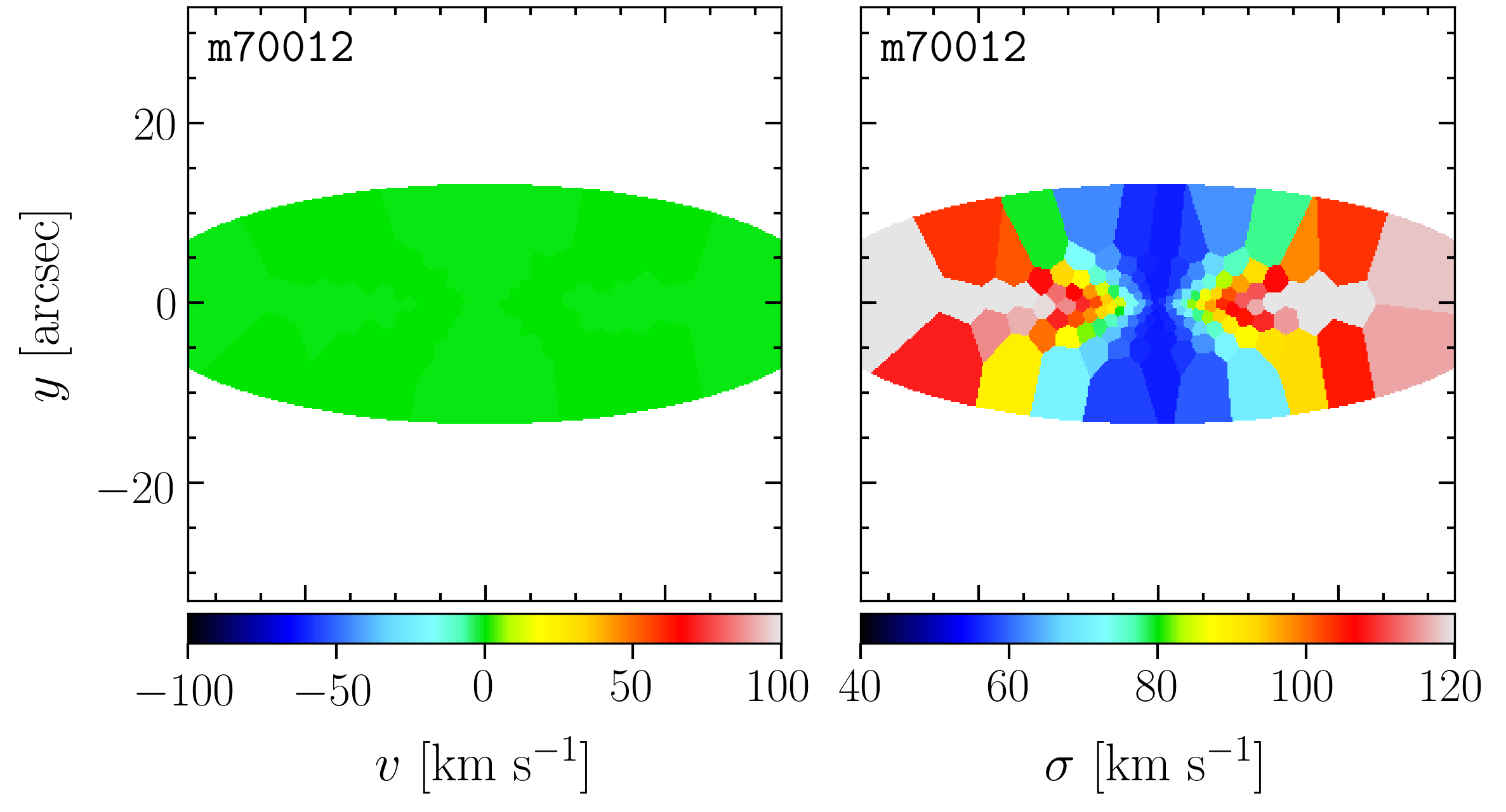}\\
\caption{
Comparison of stellar velocity and velocity dispersion maps ({\em left} and {\em right} panel of each pair of plots, respectively) of models {\tt mii004} ({\em top row\/}),
{\tt mii008} ({\em middle row\/}), and {\tt mii012} ({\em bottom row}) with the three adopted inclinations of 40$^\circ$ ({\em right column}), 50$^\circ$ ({\em middle column}), and 70$^\circ$ ({\em left column}), respectively.}
\label{fig:inclinations}
\end{figure*}

\begin{figure*}
\centering
\includegraphics[width=0.625\textwidth]{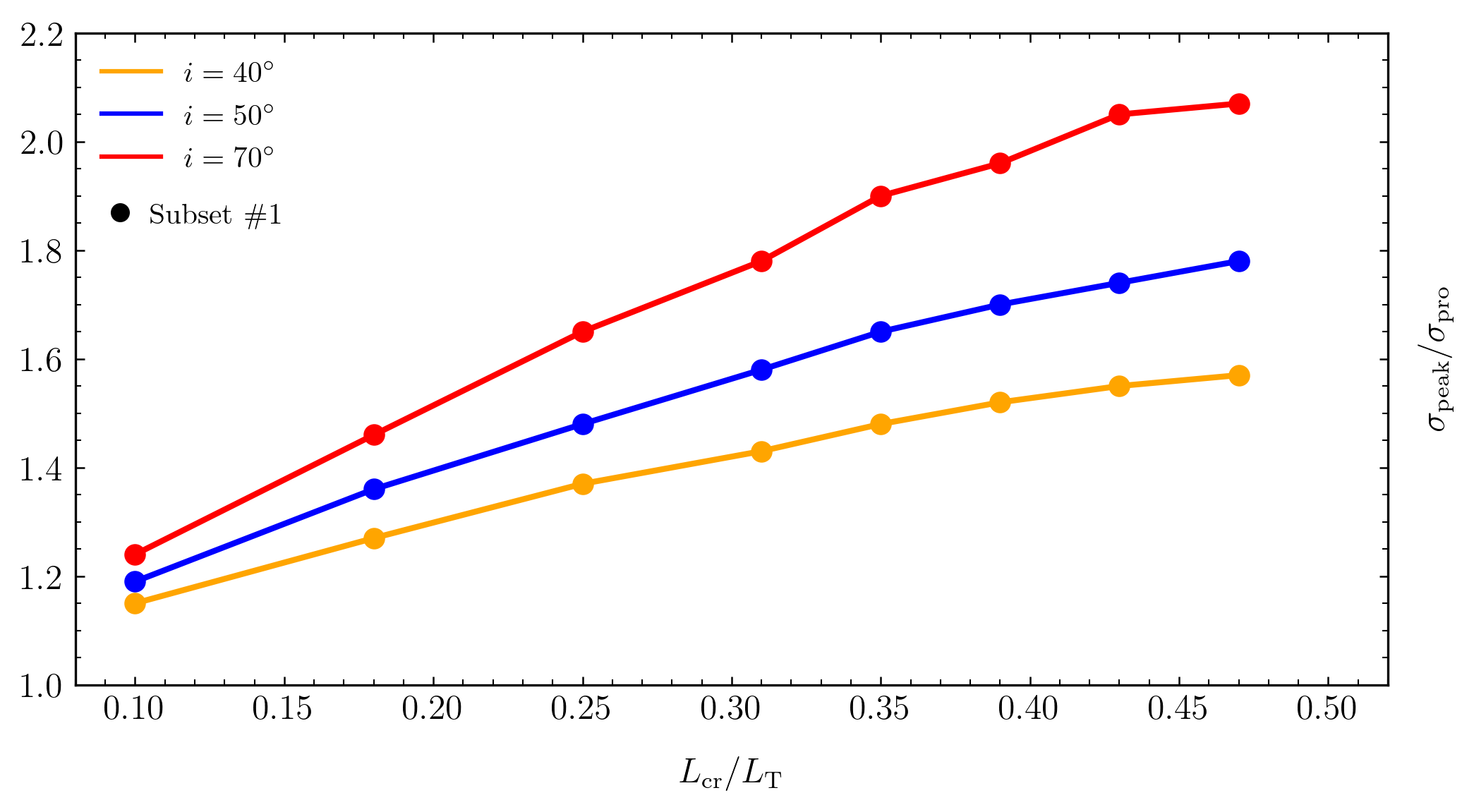}  
\caption{Ratio of the peak velocity dispersion $\sigma_{\rm peak}$ to the input velocity dispersion $\sigma_{\rm pro}$ as a function of the ratio of the total luminosity of the counter-rotating component $L_{\rm cr}$ to that of the model $L_{\rm T}$ for subset 1 ({\em filled circles\/}). The {\em yellow, blue,\/} and {\em red lines\/} show the trend for inclination $i=40^\circ, 50^\circ$, and $70^\circ$, respectively.}
\label{fig:sigmapeak}
\end{figure*}

A comparison in Fig.~\ref{fig:inclinations} of three representative models, {\tt mii004}, {\tt mii008}, and {\tt mii012}, with different inclinations shows that higher viewing angles emphasize the above features. 
Highly inclined disks cause an increase in the projected velocity separation between the two counter-rotating components. 
Thus, for example, the values of $\sigma$ at the peak locations ($r_{\rm peak}$) become larger and the slopes (from center to $r_{\rm peak}$) are steeper for higher inclinations. 
For a quantitative analysis, we extract the radial profiles along the major axis of all models with different inclinations belonging to subset 1, and measure the maximum values of $\sigma$ and their locations ($\sigma_{\rm peak}, r_{\rm peak}$). 
We find that the $\sigma$ peaks for models of subset 1 are located at radii between $\sim0.5$ and 1.5$r_{\rm e}$.
The value of $\sigma_{\rm peak}$ increases with increasing luminosity contribution of the counter-rotating disk to the model, as shown in Fig.~\ref{fig:sigmapeak}. 
From these results we also confirm that the statistics of $2\sigma$ galaxies is biased by higher viewing angles.

\subsection{Counter-rotating signatures from four-moment analysis}\label{sec:4mom}

Analyzing the second kinematic set ($v,\sigma,h_3,h_4$) we find that the mean velocity is even closer to the velocity of the brighter component compared to the previous two-moment analysis and the $2\sigma$ feature is not always present.

\begin{figure*}
\centering
\centering
\includegraphics[width=0.95\textwidth]{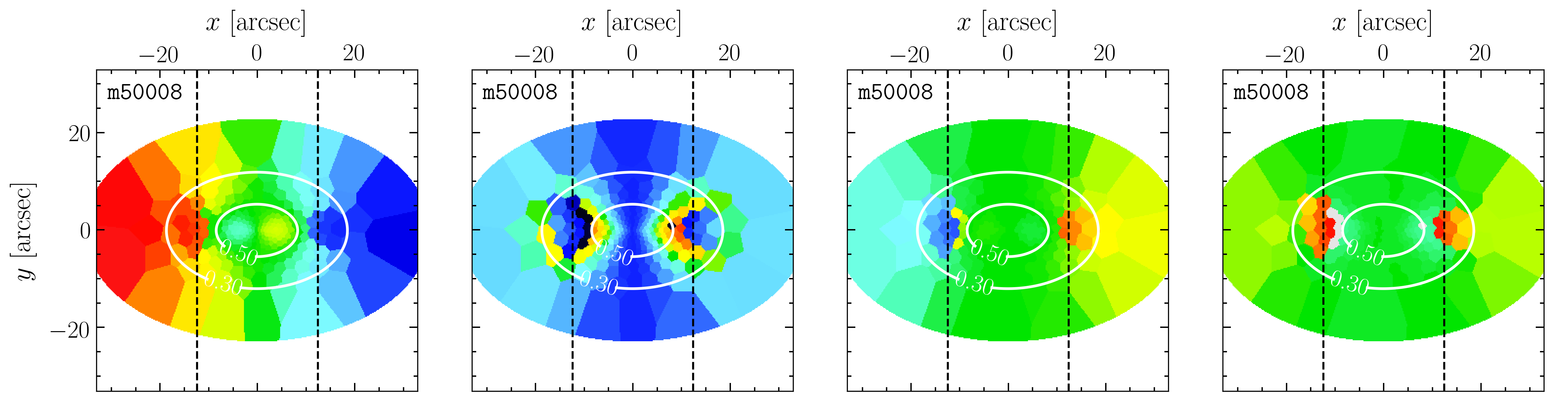}
\centering
\includegraphics[width=0.95\textwidth]{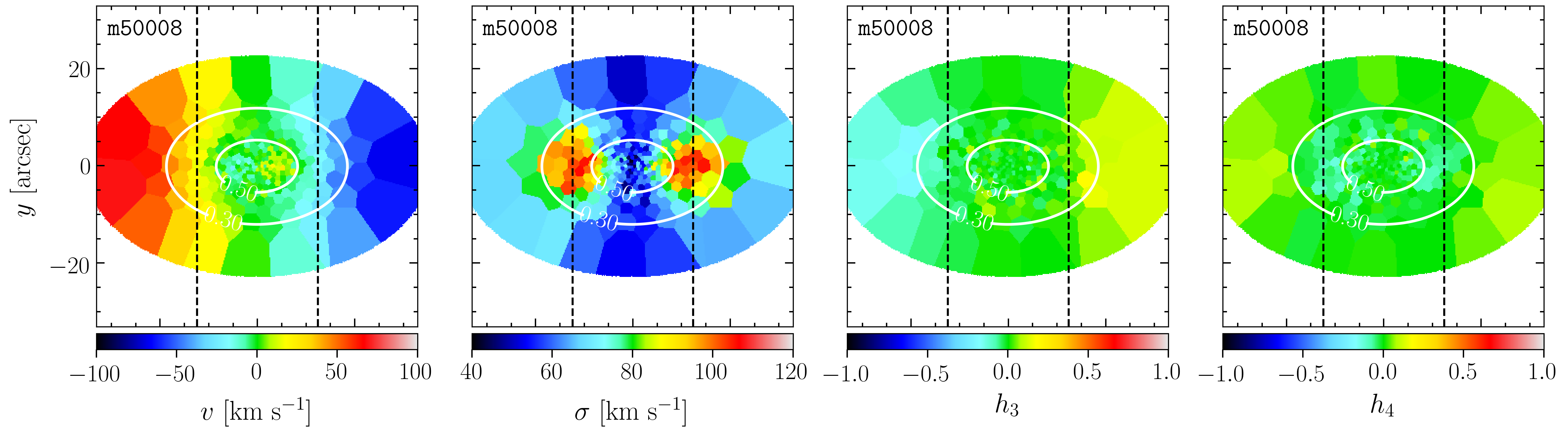}
\caption{Stellar kinematic maps of $v$, $\sigma$, $h_3$, and $h_4$ ({\em from left to right\/}) of the LOSVD for the model {\tt m50008} derived from data without noise ({\em upper panels\/}) and with $S/N=5$ pixel$^{-1}$ ({\em lower panels\/}). 
The {\em inner\/} and {\em outer dotted ellipses\/} show where $f_{\rm cr} = 0.50$ and 0.30, respectively. 
The {\em solid vertical lines\/} give the position of the $2\sigma$ drop detected along the major axis radial profile of the velocity dispersion.}
\label{fig:drop}
\end{figure*}

Instead of the {\em $2\sigma$ peak\/} we detect a new kinematic feature, which we name {\em $2\sigma$ drop\/}. 
It corresponds to the presence of two minima in the velocity dispersion profile along the major axis with 
$\sigma_{\rm drop} \le 0.9 \sigma_{\rm pro}$ at $0.5r_{\rm e}\lesssim |r|\lesssim1.5 r_{\rm e}$. 
This feature is seen only in galaxy models having $f_{\rm cr}> 0.30$ and $\Delta v > 100$ km s$^{-1}$.
Recovering a $2\sigma$ drop strongly depends on the data quality. 
The $2\sigma$ drop disappears in favor  of a $2\sigma$ peak by adding noise to the kinematic model.
In Fig.~\ref{fig:drop} we show that by adding noise to the model {\tt m50008} to have $S/N=5$ pixel$^{-1}$, we do not measure $\sigma$ lower than the input value, which results in an unphysical  solution.
This behavior allows us to clarify that we do not measure a real drop in the velocity dispersion at the center of the galaxy. 
It is an artifact due to adopted GH parameterization of the LOSVD rather than the signature of a peculiar orbital structure of the galaxy. 
On the contrary, the presence of young stars distributed in a thin and kinematically cold disk in the innermost regions of a galaxy may dominate the light distribution causing a real {$\sigma$ drop} in the very center of the velocity dispersion map \citep{portaluriKinematicsSigmadropBulges2017}.

Compared to the previous analysis, the change in $v$ and $\sigma$ values are related to the use of higher-order moments for the extraction of the LOVSD, which are worth  analyzing as well.
First of all, we recall that we assume both the prograde and counter-rotating disks to be thin stellar structures with a Gaussian LOSVD, and therefore with null $h_3$ and $h_4$ moments. 
Thus, we do not have a low-velocity tail producing asymmetric deviations traced by the $h_3$ parameter in any of the velocity distributions along the LOS of the two counter-rotating components, as usually observed in galaxy disks \citep{binneyGalacticDynamics1987,krajnovicATLAS3DProjectXVII2013,vandesandeSAMIGalaxySurvey2017}. 

We show in Sect.~\ref{sec:GHparam} that large $h_3$ and $h_4$ are an indication of strongly asymmetric or even double-peaked absorption lines, specifically strong counter-rotation. 
Hence, we focus on most critical cases of galaxy models with small total light contribution (i.e., the first four models of subset 1), which show weak counter-rotating signatures because of their small values of $h_3$ and $h_4$, except for   model {\tt m70004}, which is useful as a comparison.

\begin{figure*}
\centering
\includegraphics[width=0.8\textwidth]{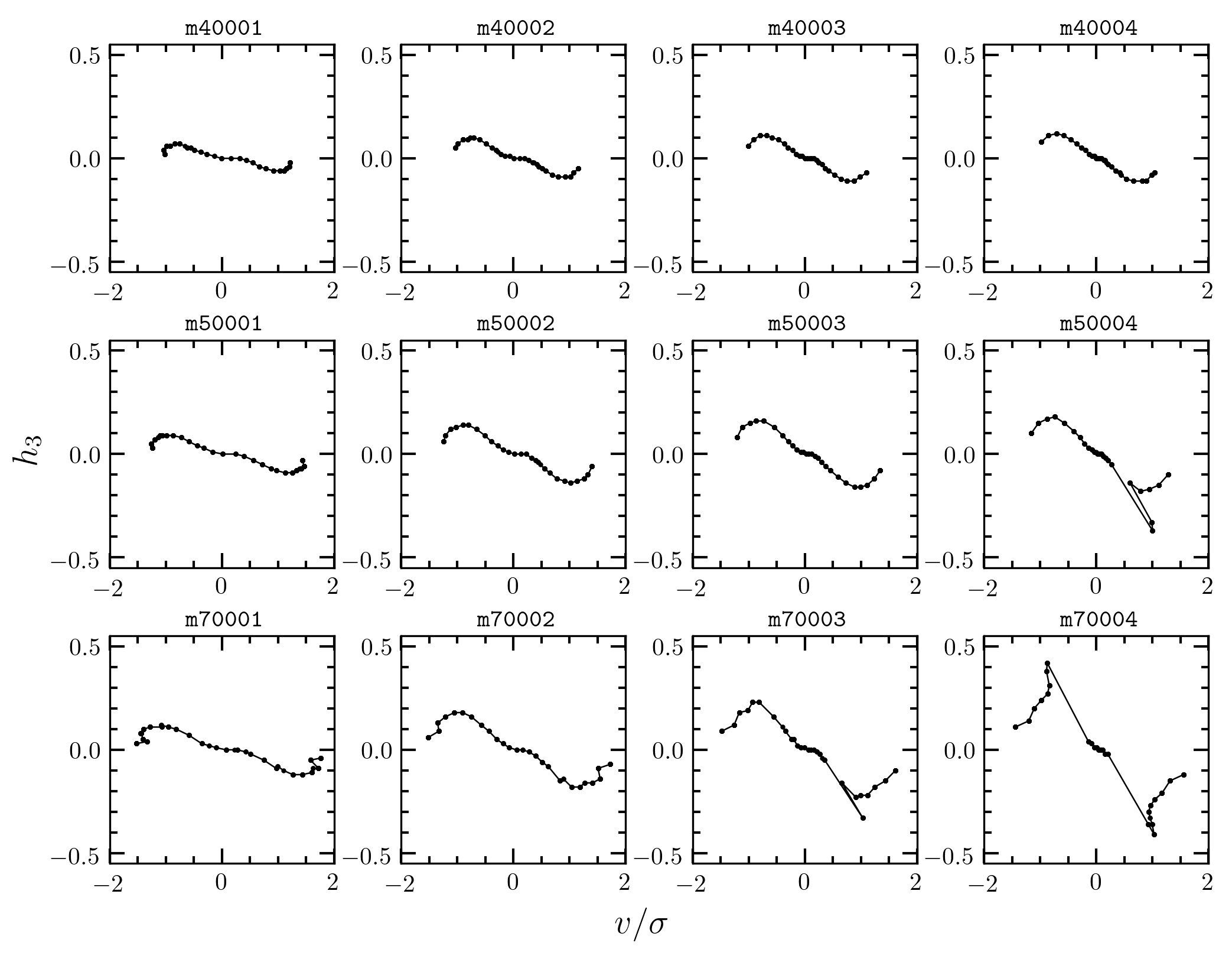}   \\
\includegraphics[width=0.8\textwidth]{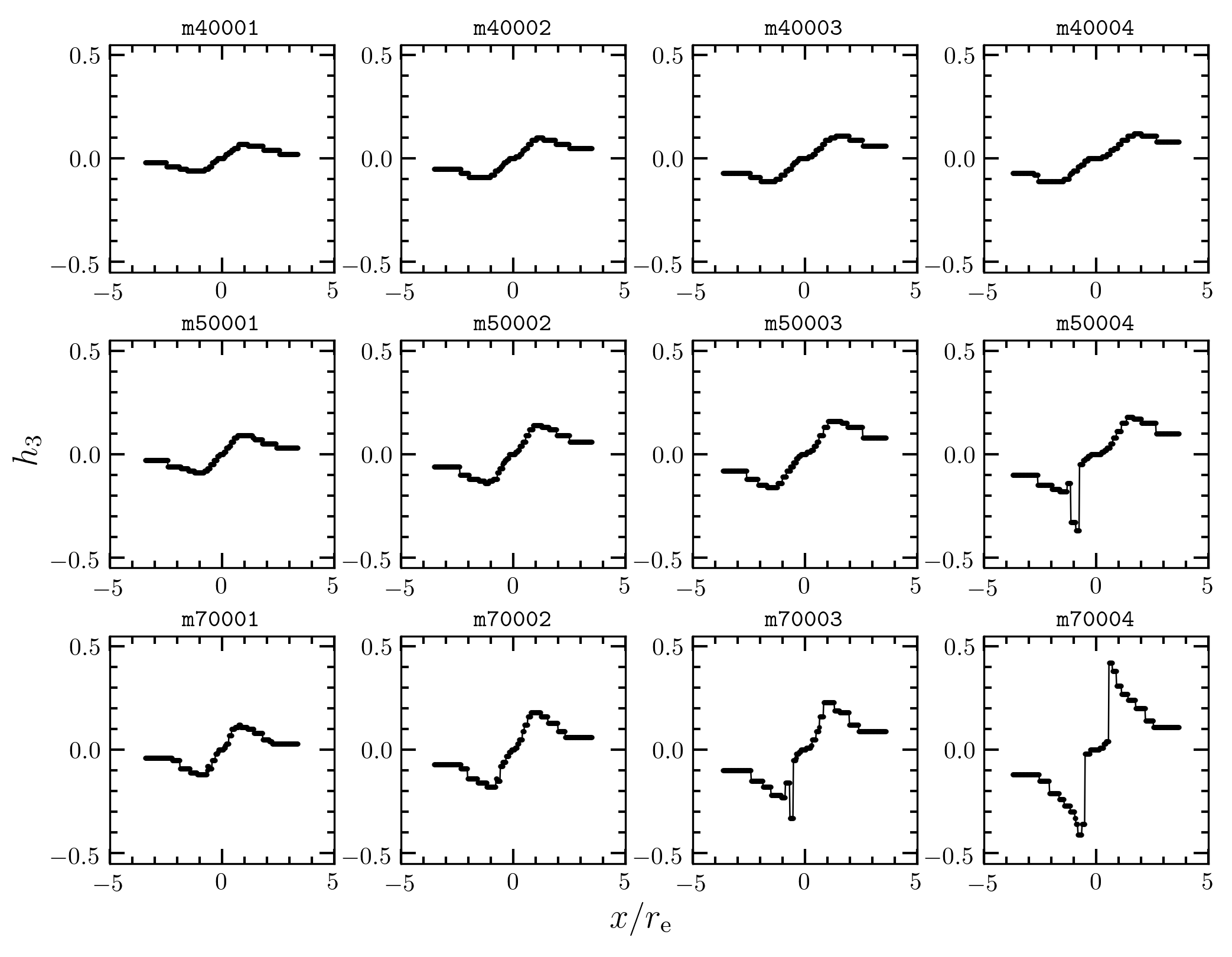}   
\caption{Local correlation of $h_3-v/\sigma$ ({\em top panels\/}) and radial trend of $h_3$ as function of $x/r_{\rm e}$ ({\em bottom panels\/}) for the first four models of subset 1 ({\em from left to right\/}) with inclination $i=40^\circ,~50^\circ$, and $70^\circ$ ({\em from top to bottom\/}, respectively). The {\em filled circles} are the data points that are connected through a {\em solid line} to better illustrate the trend.}
\label{fig:h3vs}
\end{figure*}

Our models are characterized by a local anti-correlation of $h_3$ with $v/\sigma$ where $|v/\sigma| \lessapprox 1$, as we see in Fig.~\ref{fig:h3vs}. 
This anti-correlation becomes steeper for larger contribution of the counter-rotating component and stronger for larger velocity separations (or equivalently higher inclinations). 
Models {\tt m50004} and {\tt m70003} show some points at large $h_3$. 
The explanation relies on the possible degenerate solutions of the fitting method. 
Inspecting the kinematic maps of {\tt m50004}, we see   on the left side of the model that the measured $\sigma$ is smaller, $v$ is larger, and $h_3$ and $h_4$ are larger than on the other side.
This means that for the same LOSVD we have two different parameterizations leading to different values of $v/\sigma$ and $h_3$. If the solution with the $2\sigma$ drop is preferred for model {\tt m50004}, then the local correlation would resemble that of model {\tt m70004} with a discontinuity at $v/\sigma\approx1$. 
Looking at the local values of $h_3$ as a function of $x/r_{\rm e}$ in Fig.~\ref{fig:h3vs}, we see that $h_3$ shows a change in slope at around $r_{\rm e}$, being negative for $|r|>r_{\rm e}$ and positive in the inner regions.
This correlation was studied by \citet{fisherKinematicProfilesGalaxies1997} from observations of 18 S0 galaxies. 
He found a wide variety of behaviors in the local $h_3-v/\sigma$ relation. 
In particular, the anti-correlation is always seen in the inner regions, while in the outer there are three different trends including the correlation between $h_3$ and $v/\sigma$. 
The high-speed tails can be explained as features that occur primarily within the bulge dominated region of the galaxies.
In our case it is exclusively due to the presence of a secondary, non-dominant, counter-rotating component, and probably depends on the interplay of the two different exponential profiles. 
A systematic study of the higher-order parameters at large radii along the major axis for a large sample of disk galaxies, including those with bars, is necessary to better understand the $h_3$ behavior and its possible indication of counter-rotating stellar components. 

\section{Observational limits}\label{sec:limits}
The most critical situation in detecting a counter-rotating component is for the case of weak counter-rotation and when the $2\sigma$ feature is not well defined.
To better understand whether the $h_3$ parameter is a good tracer of weak counter-rotation, we analyzed the $h_3$ radial profile of a simulated galaxy without counter-rotation evolved with an N-body code (Sect.~\ref{sec:h3}). 
For the clear identification of the $2\sigma$ peak we decided to start from a very noisy model and to increase the $S/N$ until we recovered this feature (Sect.~\ref{sec:SN}).

\subsection{Thin disk approximation}\label{sec:h3}
A possible limit of our approach is the assumption of two counter-rotating stellar disks being thin. 
Counter-rotating stellar disks may have a different thickness \citep{cappellariSAURONProjectOrbital2007, coccatoSpectroscopicEvidenceDistinct2013, pizzellaEvidenceFormationYoung2018}. 
When we observe a thick stellar disk, the integration along the LOS may give rise to non-zero values for $h_3$ and $h_4$ that could affect our kinematic diagnostics preventing us from identifying the presence of a counter-rotating component. 
This is the reason why we exclude highly inclined disks in our analysis.

However, we quantify this effect by building and analyzing a mock datacube obtained with the MUSE standard setup for a simulated galaxy.
In this simulation we model the evolution of a disk galaxy with the $N$-body+smoothed particle hydrodynamic code GASOLINE \citep{wadsleyGasolineAdaptableImplementation2004}. We start with the pure gas corona$+$dark matter halo system described in \citet{roskarInsideOutGrowthFormation2008}. 
Over the first 4 Gyr, gas cools and forms stars. 
At 4 Gyr we stop star formation and let the system evolve to 6 Gyr.  At this point we rotate the stars (only) by $180^\circ$ about the $x$-axis, and restart star formation, evolving to 11 Gyr.  While this is an ad hoc method for producing a counter-rotating galaxy, nonetheless it represents a better approximation to real counter-rotating galaxies than does our thin disk approximation.

We use the same numerical parameters as in \citet[][but see also \citealt{roskarRadialMigrationDisc2012}]{roskarInsideOutGrowthFormation2008}. At the end of the simulation, the stellar component consists of 1.8M particles corresponding to a total mass of $5.7\times10^{10}$ M$_\odot$.  The counter-rotating component contributes 34$\%$ to the total mass. 

By considering only the star particles and neglecting the contribution of gas, we generate the luminosity-weighted spaxels of a datacube. 
We use the SYNTRA code \citep{portaluriKinematicsSigmadropBulges2017} that translates the photometric, kinematic, and chemical properties of the simulated galaxy assigning a SSP model to each stellar particle.
Here, the adopted SSP spectra are from the GALAXEV library computed using the isochrone synthesis code of \citet{bruzualStellarPopulationSynthesis2003}.
For each simulation time step and given galaxy orientation (i.e., $i$ and $PA$), SYNTRA allows us to build a datacube covering a FoV of $5.3\times5.3$ kpc$^2$. 

We consider mock observations of the simulated galaxy with a $PA=0^\circ$ and two viewing angles of $60^\circ$ and $80^\circ$.
For each of the two inclinations, we build two datacubes: 1) {\tt sim0-11i60/80} including all stars, with ages of 0-11 Gyr and  2) {\tt sim6-11i60/80} considering only stars with ages of 6-11 Gyr. The former includes both prograde and retrograde stars, while the latter only   prograde stars. 
The stellar counter-rotating component is younger ($t<6$ Gyr).

\begin{figure*}
\centering
\centering
\includegraphics[width=0.95\textwidth]{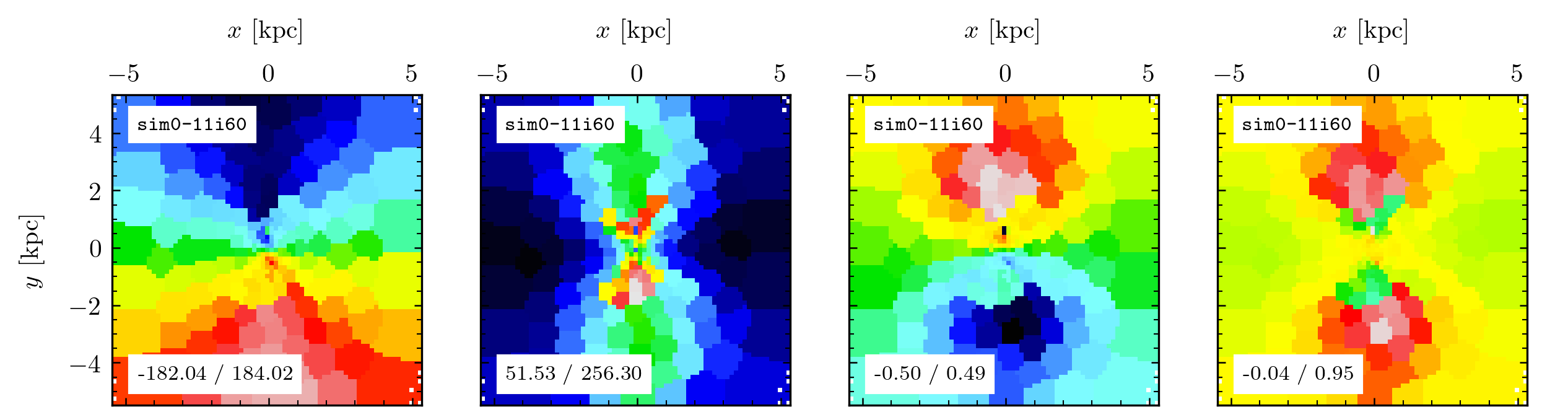}
\centering
\includegraphics[width=0.95\textwidth]{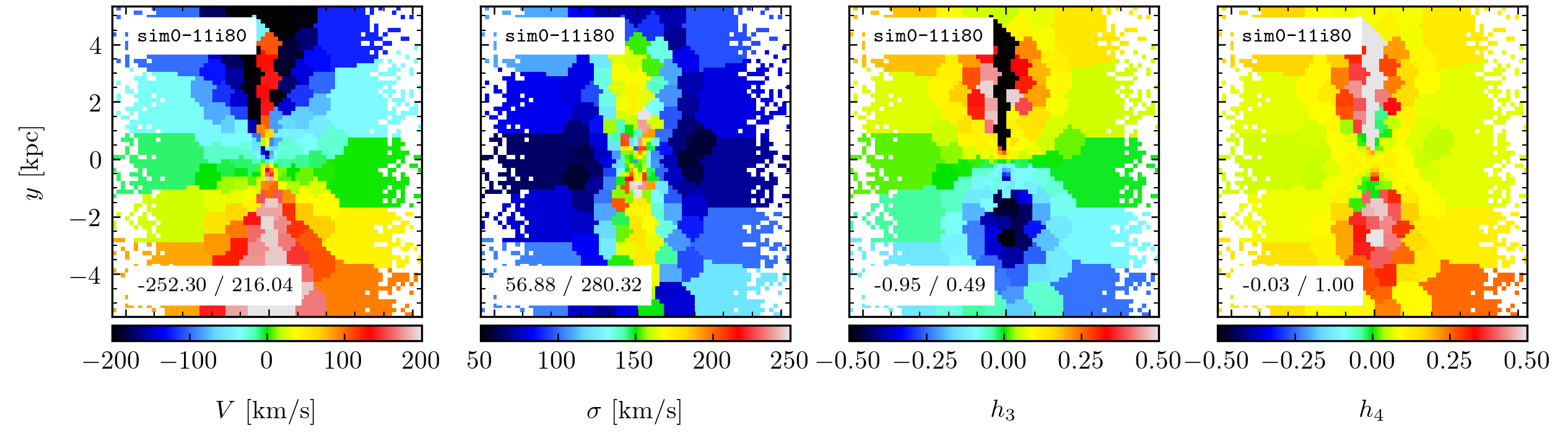}
\centering
\includegraphics[width=0.95\textwidth]{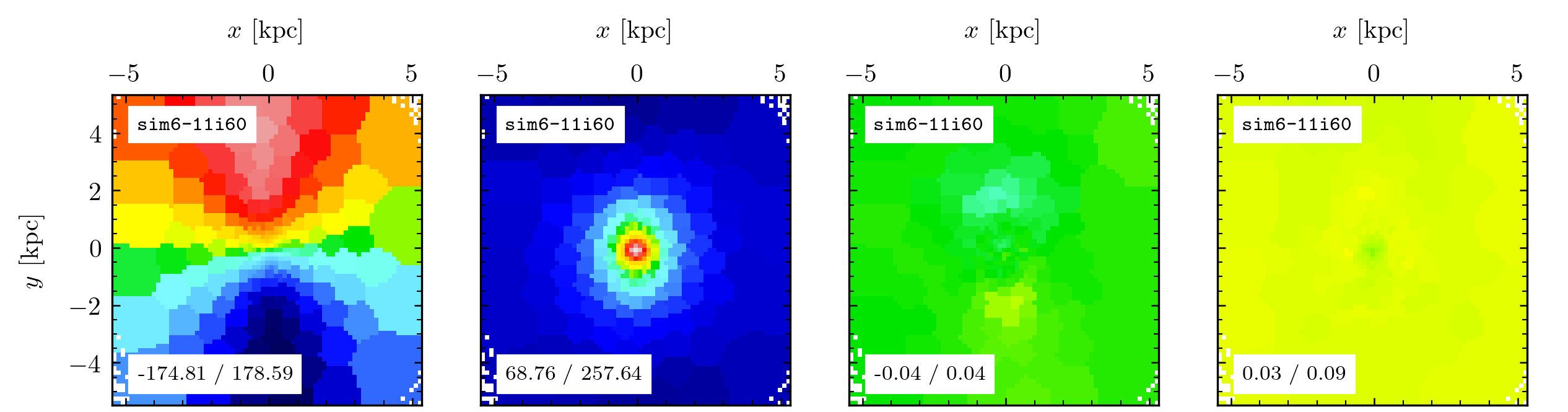}
\centering
\includegraphics[width=0.95\textwidth]{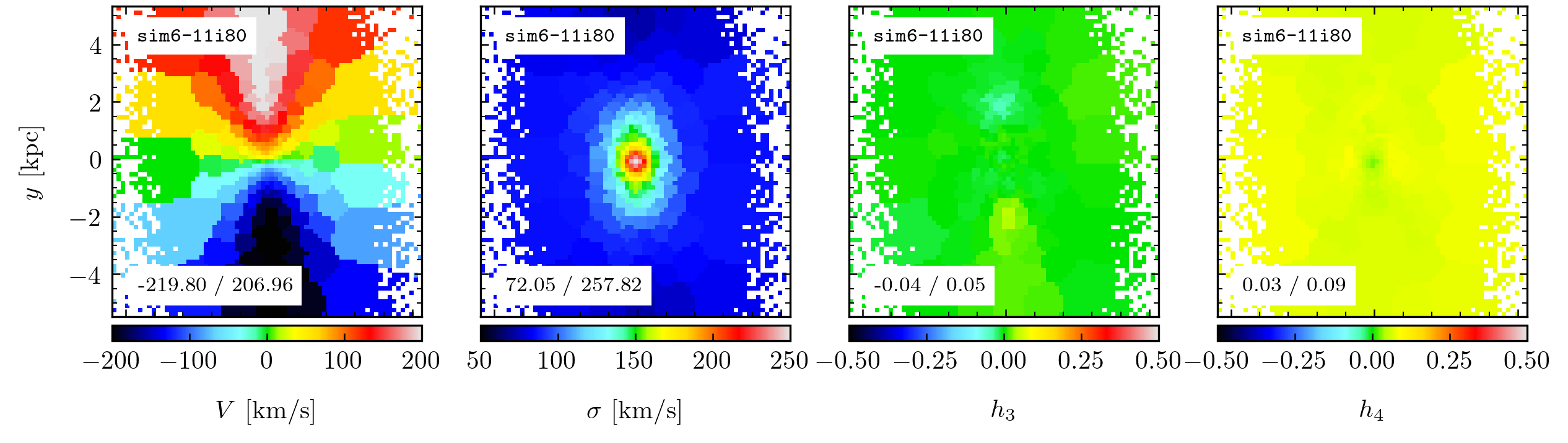}
\caption{Stellar kinematic maps of $v$, $\sigma$, $h_3$, and $h_4$ ({\em from left to right\/}) of the LOSVD for the simulated galaxy with ({\tt sim0-11i60/80}, {\em top panels\/}) and without ({\tt sim6-11i60/80}, {\em bottom panels\/}) counter-rotating stars.}
\label{fig:sim}
\end{figure*}

For the four datacubes, we measure the stellar kinematics with pPXF in the same way as we do for our kinematic models. 
In Fig.~\ref{fig:sim} we show the $v$, $\sigma$, $h_3$, and $h_4$ maps of the datacubes {\tt sim0-11i60/80} and {\tt sim6-11i60/80}.
Considering both prograde and retrograde star particles, the mock observations of the simulated galaxy further confirm the kinematic signatures of two counter-rotating stellar disks, in particular the $2\sigma$ signature, which is stronger at higher inclinations.
In the case with only a prograde disk, the $2\sigma$ signature is no longer present, while the values of $|h_3|$ only reach a maximum of 0.05. 
The local anti-correlation of $h_3$ with $v$ is limited to the bulge region, which is identified by the higher velocity dispersion at the galaxy center.
These results support the stellar counter-rotation diagnostics, which we investigated by our simple kinematic models relying on the thin disk approximation.

\subsection{The effect of the S/N}\label{sec:SN}

To investigate whether $h_3$ is a good diagnostic also in the case of weak counter-rotation when the $2\sigma$ feature is not well defined and $S/N$ is poor, we examine the galaxy model {\tt m50001} in more detail because it shows a very faint $2\sigma$ signature. 
The counter-rotating disk of {\tt m50001} is characterized by a total luminosity ratio $L_{\rm cr}/L_{\rm T}=0.11$ and a maximum surface brightness contribution $f_{\rm cr}=0.20$ in the center and decreasing outwards.
By measuring only the first two moments of the LOSVD of the the noise-free model, we find $\sigma_{\rm peak}/\sigma_{\rm pro}\sim1.2$. 
We measure a maximum value of $|h_3|=0.09$ when we also take the higher-order moments of the LOSVD into account. 

\begin{figure*}
\centering
\centering
\includegraphics[width=0.95\textwidth]{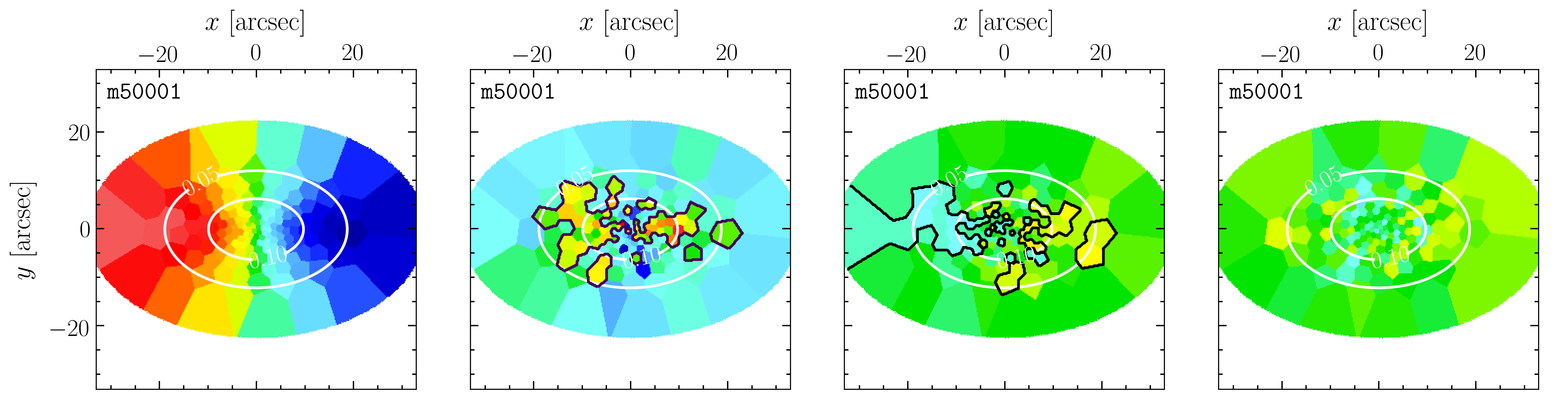}
\centering
\includegraphics[width=0.95\textwidth]{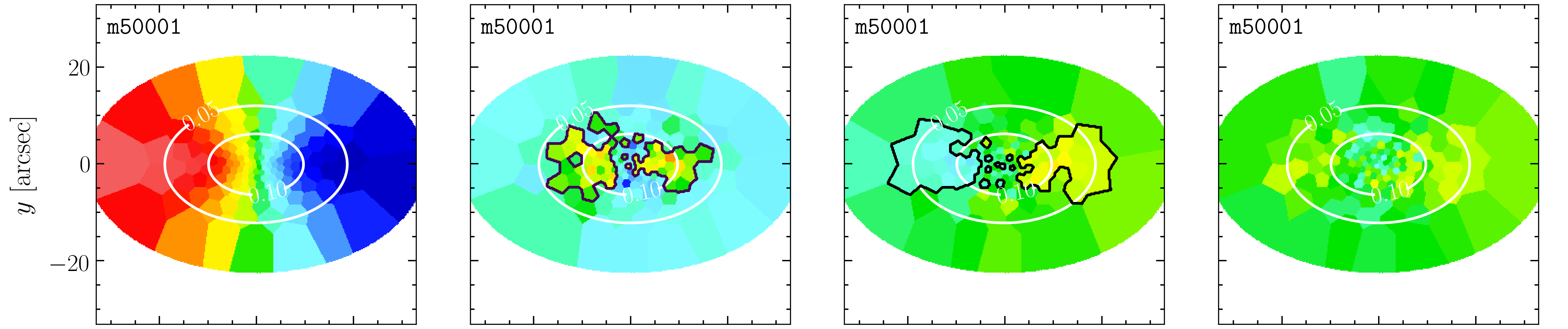}
\centering
\includegraphics[width=0.95\textwidth]{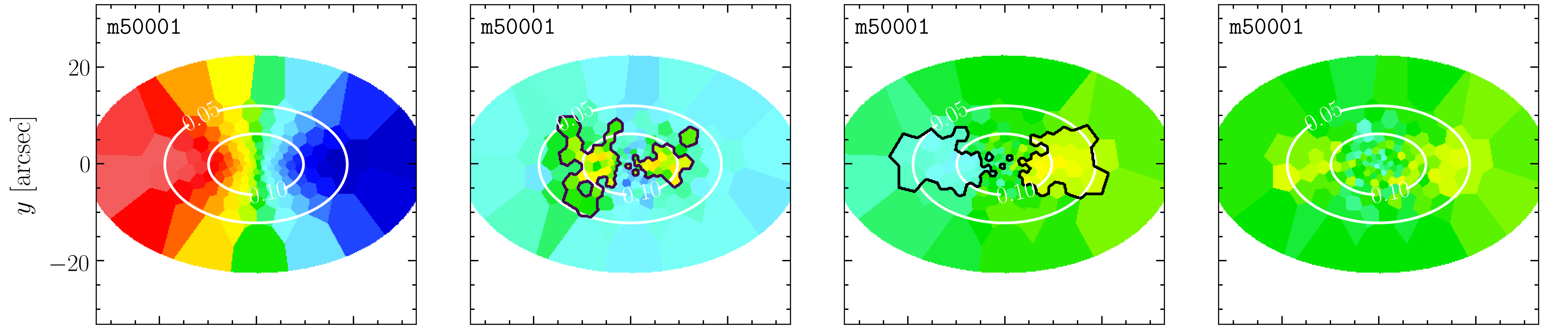}
\centering
\includegraphics[width=0.95\textwidth]{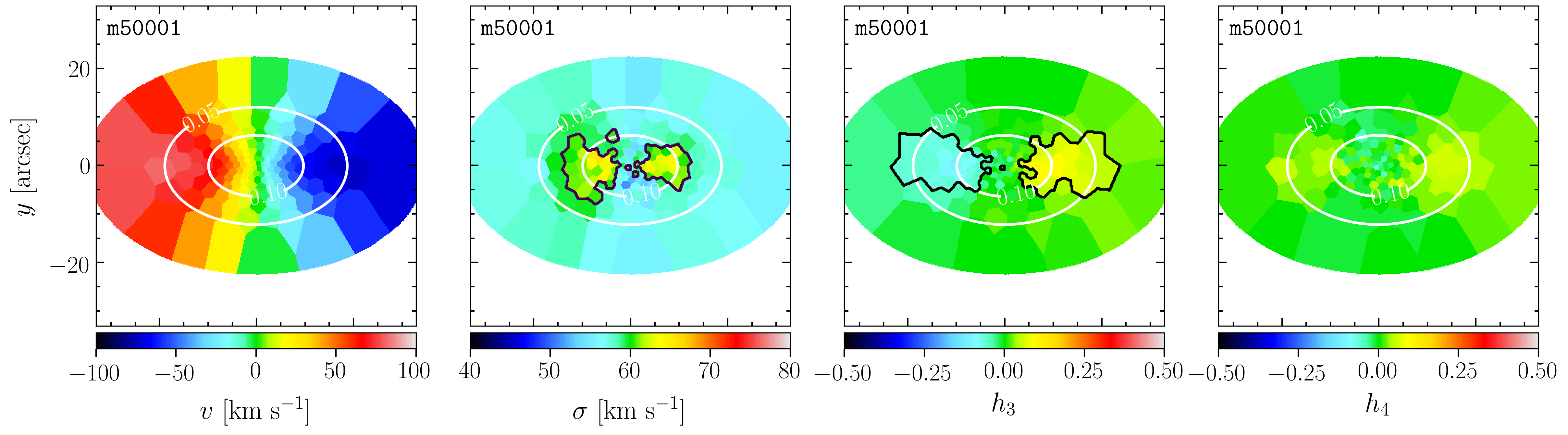}
\caption{Stellar kinematic maps of $v$, $\sigma$, $h_3$, and $h_4$ ({\em from left to right\/}) of the LOSVD for the model {\tt m50001} derived from data with 
$S/N = 5, ~10, ~15,~20$ pixel$^{-1}$ ({\em from top to bottom\/}). The {\em inner\/} and {\em outer dotted ellipses\/} show where $f_{\rm cr} = 0.10$ and 0.05, respectively. The {\em solid contours\/} in the $\sigma$ and $h_3$ maps correspond to $\sigma=60$ km s$^{-1}$ and $|h_3|=0.05$, respectively.}
\label{fig:SN}
\end{figure*}

The stellar kinematics maps we measure for {\tt m50001}, by increasing the noise to  $S/N=5, ~10,~15,$ and $20$ pixel$^{-1}$, are shown in Fig.~\ref{fig:SN}. 
We clearly identify the $2\sigma$ peak for the galaxy model with $S/N=20$ pixel$^{-1}$ when the two offset peaks of velocity dispersion are nicely outlined by the contour level at $\sigma = 60$ km s$^{-1}$, even though the increase in $\sigma$ is small relatively to $\sigma_{\rm pro}$. 
The maximum value of $|h_3|$ is $\sim0.1$ in all cases, and the $h_3$ signature extends as far as the counter-rotating component, out to surface brightness contribution $f_{\rm cr}=0.05$. 
Even at low $S/N$, the anti-correlation between $h_3$ and $v$ is more clearly discernible than the $2\sigma$ peak. We point out that our analysis considers only pure disk structures. Thus, this large-scale structure in the $h_3$ map may be misleading when other structural components are present.

In this respect, when analyzing real galaxies, where the presence of intrinsic irregular structures or inhomogeneities due to instrumental characteristics make the data analysis more challenging, an S/N treatment like the one suggested by \citet{gonzalez-gaitanSpatialFieldReconstruction2019} will   improve our results.

\section{Conclusions}\label{sec:CONC}

The signature of two large-scale counter-rotating stellar disks appears in a galaxy spectrum with a large variety of LOSVDs that are  broadened, strongly asymmetric, or even bimodal. 
This pushes to the limit the most widely used method for extracting the stellar kinematics, namely the single GH parameterization, which nevertheless provides important clues to the presence of multiple kinematic components in a galaxy.

In this work we tested the LOSVD recovery for a set of mock spectra adopting the GH parametric fit. 
Though the residuals decrease with increasing number of GH moments and $S/N$, we find that the fit to strongly non-Gaussian or bimodal spectral lines is not appropriate. 
Even allowing $|h_3|$ and $|h_4|$ to be larger than $0.3$, the best-fit solution is poor, especially when the two counter-rotating components have large velocity separation and one of them contributes less than  30$\%$ to the luminosity. 
However, the GH parameterization is useful for revealing counter-rotating signatures in spatially resolved kinematics.

The extraction of the stellar LOSVD using the GH series lead to the conclusion that by adding a counter-rotating component to a dominant prograde disk  the amplitude of the observed rotation curve is lowered, while  the observed velocity dispersion is increased. In particular, our analysis confirms that the stronger evidence of the presence of two counter-rotating stellar disks is a $2\sigma$ feature along the major axis of the galaxy.
The size, shape, and slope of the 2$\sigma$ peaks strongly depend on the velocity separation and contribution of the two counter-rotating disks.
This signature is highlighted even by measuring only the lower-order moments $v$ and $\sigma$ of the GH truncated series.

Including the higher-order moments to the fit, we find a change in the observed mean velocity, which becomes closer to the velocity of the brighter component compared to that measured by fitting $v$ and $\sigma$ only. 
Concerning the velocity dispersion, we find that the $2\sigma$ feature is not always present.
It disappears in favor of a new kinematic features, called the  {$2\sigma$ drop}, which is an artifact of the adopted parameterization when fitting high $S/N$ data. 
In other words, the code detects only one component, which is the dominant one. Thus, the best-fit solutions give velocities and velocity dispersions closer to those of the single component. However, the contribution of the secondary disk is still revealed by the higher-order moments.
By adding noise to the spectra, the $2\sigma$ peak is recovered, while the higher-order signatures become less prominent.

At low $S/N$ regimes and for faint counter-rotating stellar component, the $2\sigma$ feature may be unclear.
Interestingly, we find that the local anti-correlation of $h_3$ with $v$ is always a useful indicator of counter-rotation as long as we consider only the disk structure.
In addition, the $h_3$ parameter provides a criterion to distinguish between strong and weak counter-rotation.
When the asymmetries in the LOSVD become relevant, then the $|h_3|$ assumes values larger than $0.2$.
Contrary to the $2\sigma$ feature, the large-scale $h_3$ signature weakly fades with the decrease in $S/N$. Independently of the $S/N$, all the signatures become fainter at lower viewing angles as an effect of the smaller projected velocity separation between the two counter-rotating stellar components. 
Hence, highly inclined galaxies are favored for the application of the kinematic diagnostics, limiting the statistics of counter-rotating stellar disks.

A fundamental step in improving the statistics of counter-rotating galaxies is to detect faint counter-rotating stellar disks while  looking for the $2\sigma$ feature, but also analyzing the signatures in the higher-order moment maps. 
The counter-rotation diagnostics confirmed in this work can be used as a guideline to search for counter-rotating signatures in the kinematic maps of galaxies stored in the MUSE data archive.
This research will increase the number of optimal candidates to test and fine-tune the state-of-the-art decomposition techniques, in particular those based on the dynamical modeling using kinematic constrains.
The study of large-scale counter-rotating galaxies will help to determine the relative importance of the different formation scenarios probing the galaxy assembly models in the cosmological context.

Present-day research on extended counter-rotation is limited by the small number of objects known to host this phenomenon. 
The IFU data available in the archives still do not have the necessary spatial extent and spatial resolution to perform complete statistics. 
In the near future, the IFU mounted on next-generation
telescopes will allow us to investigate counter-rotators or, more  generally,
decoupled components \citep{corsiniPolarBulgesPolar2012}, at large distances and therefore at an earlier
stage of their evolution. 
For instance, the MCAO Assisted Visible Imager and Spectrograph (MAVIS) instrument at the Very Large Telescope \citep{mcdermidPhaseScienceCase2020} with a 50~mas spaxels (FoV of $5\times7.2$ arcsec$^2$) and the low-resolution spectrograph will allow us to investigate counter-rotation in galaxies at $z=0.4-0.6$. The High Angular Resolution Monolithic Optical and Near-infrared Integral field spectrograph (HARMONI) at the European-Extremely Large Telescope will also provide a similar configuration.

\begin{acknowledgements}
We thank the anonymous referee for the constructive report that helped us to improve the paper.
We thank A. Moiseev and W. W. Zeilinger for their valuable comments.
MR is grateful to P. Jethwa, G. van de Ven, and J. Falc\'{o}n-Barroso for inspiring discussions over the course of this project, and thanks the Department of Astrophysics, University of Vienna, for hospitality during the preparation of this paper.
MR, AP, EMC, and EDB are supported by MIUR grant PRIN 2017 20173ML3WW\_001 and Padua University grants DOR1885254/18, DOR1935272/19, and DOR2013080/20.
VPD is supported by STFC Consolidated grant ST/R000786/1.

\end{acknowledgements}

\bibliographystyle{aa} 
\bibliography{biblio}


\end{document}